\documentclass[sigconf, nonacm]{acmart}
\newcommand\vldbdoi{XX.XX/XXX.XX}

\newcommand\vldbpagestyle{plain} 

\usepackage{algorithm}
\usepackage[noend]{algpseudocode}

\usepackage{color}
\usepackage{enumitem}
\usepackage{multirow}
\usepackage{tabularx}
\usepackage{graphicx}

\usepackage{makecell}
\usepackage{subcaption}
\usepackage{caption}
\usepackage{multicol}
\usepackage{stfloats}
\usepackage{booktabs}
\usepackage{adjustbox}
\usepackage{tcolorbox}

\usepackage[dvipsnames]{xcolor}
\usepackage{colortbl}

\usepackage{tikz}

\newtheoremstyle{bolddef}
{}                % Space above
{}                % Space below
{\itshape}                % Body font
{}                % Indent amount
{\bfseries\scshape}       % Theorem head font
{}               % Punctuation after theorem head
{1em}               % Space after theorem head
{}                % Theorem head spec

\newtheoremstyle{exampledef}
{}                % Space above
{}                % Space below
{\itshape}                % Body font
{\parindent}                % Indent amount
{\scshape}       % Theorem head font
{.}               % Punctuation after theorem head
{.5em}               % Space after theorem head
{}                % Theorem head spec

\theoremstyle{bolddef}

\theoremstyle{exampledef}

\usepackage{listings}
\definecolor{dkgreen}{rgb}{0,0.6,0}
\definecolor{dkred}{rgb}{0.6,0,0}

\setlist[itemize]{leftmargin=10pt}
\begin{document}
\title{SVFusion: A CPU-GPU Co-Processing Architecture for Large-Scale Real-Time Vector Search}
% Real-Time Update for Large-Scale Vector Search via CPU-GPU Co-Processing
% \setlength{\belowcaptionskip}{-0.05cm}
%%
%% The "author" command and its associated commands are used to define the authors and their affiliations.

% \author{Yuchen Peng$^{1}$, Dingyu Yang$^{1}$, Zhongle Xie$^{1}$, Ji Sun$^{2}$, Lidan Shou$^{1}$, Ke Chen$^{1}$, Gang Chen$^{1}$}
% \affiliation{
% \institution{$~^{1}$Zhejiang University,  $~^{2}$Huawei Technologies Co., Ltd}
% }
% \email{{zjupengyc, yangdingyu, xiezl, should, chenk, cg}@zju.edu.cn, sunji11@huawei.com}

\author{Yuchen Peng}
\affiliation{%
  \institution{Zhejiang University$^\dag$}
}
\affiliation{%
  \institution{Hangzhou High-Tech Zone (Binjiang) Institute of Blockchain and Data Security}
}
\email{zjupengyc@zju.edu.cn}

\author{Dingyu Yang}
\authornote{Corresponding authors: Dingyu Yang, Lidan Shou}
\affiliation{%
  \institution{Zhejiang University$^\dag$\authornote{The State Key Laboratory of Blockchain and Data Security}}
}
\affiliation{%
  \institution{Hangzhou High-Tech Zone (Binjiang) Institute of Blockchain and Data Security}
}
\email{yangdingyu@zju.edu.cn}

\author{Zhongle Xie}
\affiliation{%
  \institution{Zhejiang University$^\dag$}
}
\affiliation{%
  \institution{Hangzhou High-Tech Zone (Binjiang) Institute of Blockchain and Data Security}
}
\email{xiezl@zju.edu.cn}

\author{Ji Sun}
\affiliation{%
  \institution{Huawei Technologies Co., Ltd}
  \department{}
  \streetaddress{}
  \city{}
  \state{}
  \postcode{}
}
\email{sunji11@huawei.com}

\author{Lidan Shou}
\authornotemark[1]
\affiliation{%
  \institution{Zhejiang University$^\dag$}
}
\affiliation{%
  \institution{Hangzhou High-Tech Zone (Binjiang) Institute of Blockchain and Data Security}
}
\email{should@zju.edu.cn}

\author{Ke Chen}
\affiliation{%
  \institution{Zhejiang University$^\dag$}
}
\affiliation{%
  \institution{Hangzhou High-Tech Zone (Binjiang) Institute of Blockchain and Data Security}
}
\email{chenk@zju.edu.cn}

\author{Gang Chen}
\affiliation{%
  \institution{Zhejiang University$^\dag$}
}
\affiliation{%
  \institution{Hangzhou High-Tech Zone (Binjiang) Institute of Blockchain and Data Security}
}
\email{cg@zju.edu.cn}

\begin{abstract}
Approximate Nearest Neighbor Search (ANNS) underpins modern applications such as information retrieval and recommendation. With the rapid growth of vector data, efficient indexing for real-time vector search has become rudimentary.
Existing CPU-based solutions support updates but suffer from low throughput, while GPU-accelerated systems deliver high performance but face challenges with dynamic updates and limited GPU memory, resulting in a critical performance gap for continuous, large-scale vector search requiring both accuracy and speed.
In this paper, we present SVFusion, a GPU-CPU-disk collaborative framework for real-time vector search that bridges sophisticated GPU computation with online updates.
SVFusion leverages a hierarchical vector index architecture that employs CPU-GPU co-processing, along with a workload-aware vector caching mechanism to maximize the efficiency of limited GPU memory.
It further enhances performance through real-time coordination with CUDA multi-stream optimization and adaptive resource management, along with concurrency control that ensures data consistency under interleaved queries and updates.
Empirical results demonstrate that SVFusion achieves significant improvements in query latency and throughput, exhibiting a 20.9$\times$ higher throughput on average and 1.3$\times$ to 50.7$\times$ lower latency compared to baseline methods, while maintaining high recall for large-scale datasets under various streaming workloads.
\end{abstract}

\maketitle
\thispagestyle{plain}
\pagestyle{\vldbpagestyle}

% \begingroup\small\noindent\raggedright\textbf{PVLDB Reference Format:}\\
% \vldbauthors. \vldbtitle. PVLDB, \vldbvolume(\vldbissue): \vldbpages, \vldbyear.\\
% \href{https://doi.org/\vldbdoi}{doi:\vldbdoi}
% \endgroup
% \begingroup
% \renewcommand\thefootnote{}\footnote{\noindent
% This work is licensed under the Creative Commons BY-NC-ND 4.0 International License. Visit \url{https://creativecommons.org/licenses/by-nc-nd/4.0/} to view a copy of this license. For any use beyond those covered by this license, obtain permission by emailing \href{mailto:info@vldb.org}{info@vldb.org}. Copyright is held by the owner/author(s). Publication rights licensed to the VLDB Endowment. \\
% \raggedright Proceedings of the VLDB Endowment, Vol. \vldbvolume, No. \vldbissue\ %
% ISSN 2150-8097. \\
% \href{https://doi.org/\vldbdoi}{doi:\vldbdoi} \\
% }\addtocounter{footnote}{-1}\endgroup

% \ifdefempty{\vldbavailabilityurl}{}{
% \vspace{.3cm}
% \begingroup\small\noindent\raggedright\textbf{PVLDB Artifact Availability:}\\
% The source code, data, and/or other artifacts have been made available at \url{https://github.com/zjuDBSystems/svfusion}.
% \endgroup

\section{Introduction}
\label{sec:intro}
The rise of deep learning (DL) and large language models (LLMs) has driven the widespread use of high-dimensional vector embeddings for unstructured data such as text, images, and audio.
Efficient retrieval of these embedding vectors is critical for tasks like personalized recommendation~\cite{facebook-search}, web search~\cite{bing-search}, and LLM-based applications~\cite{RAG}. 
To meet strict low-latency and high-accuracy requirements under service-level objectives (SLOs), modern vector search systems increasingly employ Approximate Nearest Neighbor Search (ANNS)~\cite{liu2004investigation}, specifically graph-based solutions~\cite{hnsw, freshdiskann, diskann, NSG, elpis} that balance accuracy and computational efficiency in high-dimensional spaces.
Building on this, recent work explores GPU-accelerated optimizations, including SONG~\cite{SONG}, GANNS~\cite{GANNS}, and CAGRA~\cite{CAGRA}, leveraging massive parallelism and high memory bandwidth to achieve significant speedups over CPU-based approaches, enabling scalable high-throughput vector search.

\begin{figure}[t]
\centering
\includegraphics[width=\linewidth]{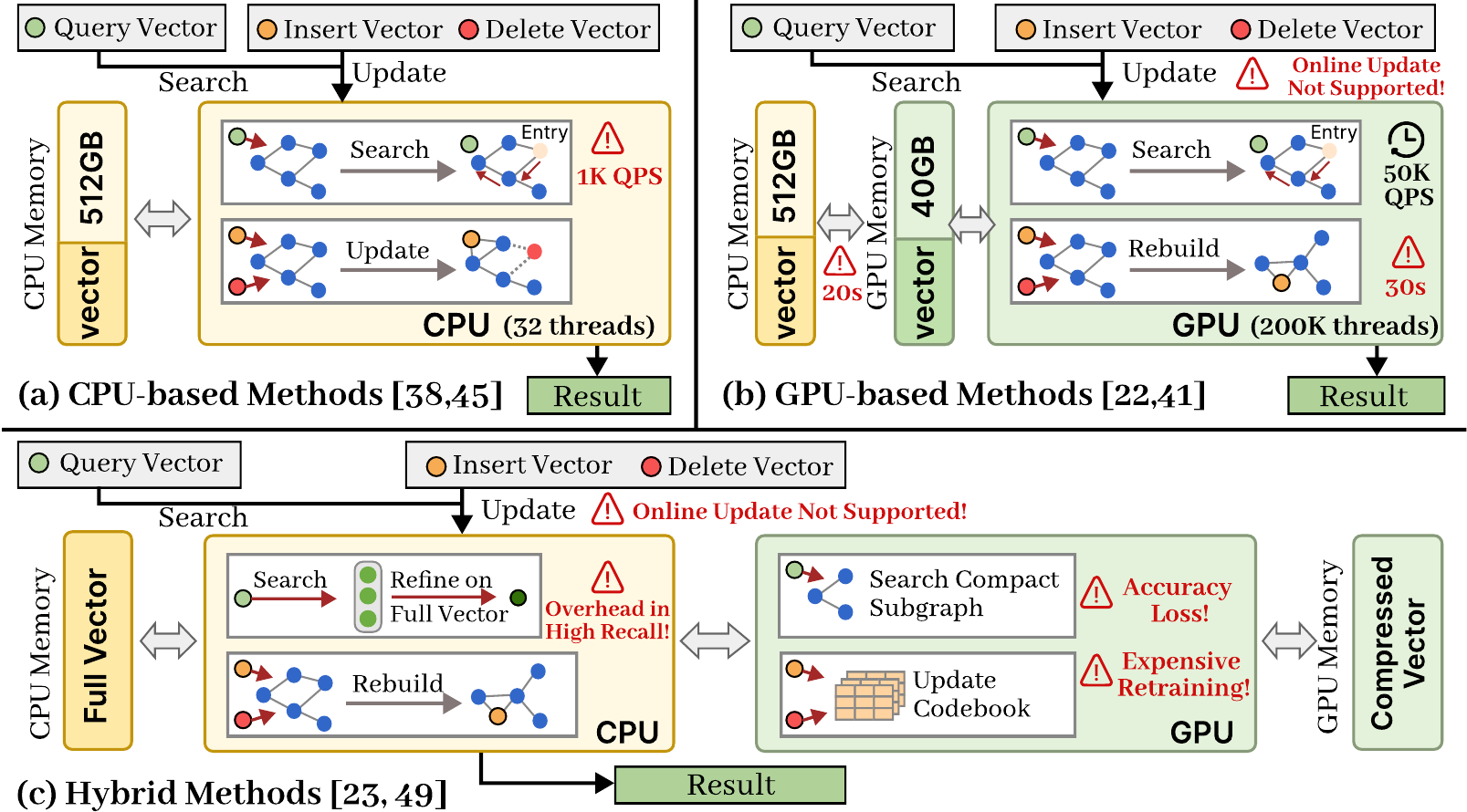}
\vspace{-2em}
\caption{Comparison of existing methods for ANNS.}
\label{fig:intro_fig}
\vspace{-1.5em}
\end{figure}

Meanwhile, the explosive growth of continuously generated embeddings has led to a fundamental shift in vector search from offline indexing to streaming processing~\cite{letta, MemGPT, RAGCache, Vearch, Gemini}. 
In real-time use cases like emergency response~\cite{streaming_rag}, retrieval-augmented generation (RAG) systems continuously ingest live data while retrieval relies upon previous results, requiring index freshness to maintain coherent LLM responses.
E-commerce platforms such as JD.com~\cite{Vearch} serve millions of independent search queries that must reflect inventory changes.
In live-video analytics and surveillance, streams yield vast numbers of embedding vectors per second, which must be both continuously indexed and queried for tasks such as object tracking and anomaly detection~\cite{rong2021scheduling}.
These applications demand not only real-time data synchronization under continuous updates but also higher processing throughput and lower latency.

Despite significant advances in high-performance vector search, most existing systems are designed for static or slowly evolving datasets, making them ill-suited for dynamic environments.
As illustrated in Figure~\ref{fig:intro_fig}, large-scale streaming scenarios demand support for search, insertion, and deletion operations~\cite{AlloyDB}.
CPU-based methods~\cite{freshdiskann,hnsw,SPFresh} offer real-time update capabilities but suffer from high computational overhead and limited memory bandwidth, resulting in poor scalability as datasets grow.
Conversely, GPU-based methods~\cite{GGNN,CAGRA} achieve high throughput for static indices, but incur significant transfer overhead when handling large-scale updates~\cite{BANG}. Moreover, complete index reconstruction is impractical for high-velocity streaming workloads.

The discrepancy between CPU-based systems (update-friendly but performance-limited) and GPU-based systems (high-throughput but static) motivates a unified hybrid approach.
Recent methods~\cite{FusionANNS, PilotANN} attempt to bridge this gap by storing compressed vectors in GPU memory, but they introduce accuracy degradation unsuitable for high-recall requirements and rely on static quantization codebooks requiring expensive retraining to handle streaming updates~\cite{RaBitQ}.
To summarize, there exist two challenges in tackling large-scale streaming ANNS:

% 大规模的向量数据如何快速搜索问题，比如diskann等，我觉得一个从数据规模出发去讲，规模大于内存或者显存量的问题如何处理，还存在什么问题
\textbf{Challenge 1: GPU Memory Capacity Constraints.} 
Modern vector data applications handle large-scale datasets with millions to billions of high-dimensional embeddings~\cite{manu, OpenaiEmbedding}, often exceeding GPU memory capacity, resulting in poor performance.
For example, serving 35M vectors in 768 dimensions~\cite{wikipedia} requires at least 120GB, far surpassing the typical 40GB capacity of high-end accelerators like NVIDIA A100. 
This limitation necessitates frequent data movement between GPU and host memory, where the performance benefits of GPU parallelism are severely diminished by costly transfer overheads~\cite{StonebrakerP24}.
While compression techniques~\cite{PQ,FusionANNS} can reduce storage requirements, they typically degrade search accuracy or increase latency to maintain comparable results.
This motivates the need for efficient CPU-GPU collaborative frameworks that effectively manage data placement.

% \textbf{Challenge 2: Performance Degradation Under Frequent Vector Updates.}
\textbf{Challenge 2: Performance Degradation Under Frequent Vector Updates.}
High-frequency vector updates at scale pose two critical performance issues.
First, existing graph-based approaches suffer from limited \textit{update throughput} during frequent insertions and deletions~\cite{hnsw,diskann}, often relying on synchronous delete marking and periodic graph rebuilds, which cause latency spikes and hinder responsiveness.
On GPU, this problem is exacerbated by frequent kernel synchronizations required for graph topology modifications~\cite{RTAMS-GANNS}, along with additional overhead from cross-device data transfers in hybrid CPU-GPU systems~\cite{RTAMS-GANNS,FusionANNS}.
Second, \textit{search accuracy} degrades over time as frequent updates deteriorate graph structure, leading to suboptimal traversal paths~\cite{CAGRA, CagraIssue}. 
% This accuracy degradation stems from the incompatibility between CPU-oriented update strategies and GPU-optimized graph structures, which disrupts connectivity and reduces recall~\cite{CAGRA, CagraIssue}.

Regarding these challenges, a question arises: can we develop a hybrid approach that avoids data compression while efficiently supporting high-frequency vector updates?
In this work, we propose SVFusion, a CPU-GPU cooperative framework for large-scale streaming vector search that bridges GPU acceleration with real-time updates to achieve high throughput while maintaining low latency.
SVFusion is built upon three key innovations:
(1) \textbf{Hierarchical Vector Index Structure} that organizes vector data and graph topology across GPU and CPU memory and disk to maximize computational efficiency.
We design a real-time coordination mechanism with
CUDA multi-stream optimization and adaptive resource management to achieve both high throughput and low latency.
(2) \textbf{Workload-Aware Vector Placement Mechanism} that adaptively determines whether to transfer vectors between CPU and GPU, or to perform computations locally on CPU when the required vectors are not available in GPU memory.
By evaluating both access patterns and graph structure, this mechanism strikes a better trade-off between data locality and transfer cost, delivering high efficiency and performance stability under dynamic workloads.
(3) \textbf{Concurrency Control for Real-time Updates} that enables safe and high-performance concurrent queries and updates.
We provide a consistency guarantee combining fine-grained locking with cross-tier synchronization, and a multi-version mechanism to ensure consistency and superior performance. 

Building on these techniques, SVFusion achieves high-performance real-time vector search with update capabilities.
In summary, we make the following contributions:
\begin{itemize}
    \item We present \textit{SVFusion}, a novel GPU-CPU-disk cooperative framework for streaming vector search that integrates hierarchical indexing with real-time coordination to maximize computational efficiency across heterogeneous memory tiers. (\S\ref{sec:overview}).
    \item We design a workload-aware placement mechanism that efficiently balances data caching and computation decisions, optimizing performance under dynamic workloads. (\S\ref{sec:system_design})
    \item We propose a concurrency control protocol that maintains data consistency through fine-grained locking and multi-version synchronization techniques. (\S\ref{sec:index_update})
    \item We implement \textit{SVFusion} and conduct a comprehensive evaluation under diverse streaming workloads to verify its effectiveness and efficiency. Results show that our framework achieves up to 9.5$\times$ higher search throughput, 71.8$\times$ higher insertion throughput, and $1.3\times$ to $50.7\times$ lower latency while maintaining superior recall rates compared to baseline methods. 
    (\S\ref{sec:evaluation})
\end{itemize} 
\section{Preliminary}
\label{sec:background}
We first define the Approximate Nearest Neighbor Search (ANNS) problem and its streaming variant (SANNS), followed by an overview of the graph-based indexing techniques. 
Table~\ref{tab:notations} summarizes the frequently used notation.

\subsection{Approximate Nearest Neighbor Search}
Let $X = \{x_1, x_2, \ldots, x_N\} \subset \mathbb{R}^D$ denote a dataset consisting of $N$ vectors, where each data point $x_i\in \mathbb{R}^D$ represents a $D$-dimensional real-valued vector in $\mathbb{R}^D$. 
The distance between any two vectors $u,v \in\mathbb{R}^D$ is denoted as $dist(u,v)$ with the Euclidean distance (i.e., the L2 norm). Given a query vector $q \in \mathbb{R}^D$ and a positive integer $k$ ($0 < k < N$), the \emph{nearest neighbor search} (NNS) problem~\cite{nns} aims to identify a subset $U_{\text{NNS}} \subset X$ containing the $k$ vectors that are closest to $q$, satisfying the condition that for any $x_i \in U_{\text{NNS}}$ and any $x_j \in X \setminus U_{\text{NNS}}$, it holds that $\text{dist}(x_i, q) \leq \text{dist}(x_j, q)$.

\begin{table}[t]
\centering
\footnotesize
\caption{Summary of notations.}
\label{tab:notations}
\vspace{-1.5em}
\begin{tabular}{|l|p{0.7\columnwidth}|}
\hline
\textbf{Notation} & \textbf{Definition} \\
\hline
$X$ & the set of $N$ $D$-dimensional vectors \\
$q$ & a query vector \\
$k$ & the number of returned results in ANNS \\
$G = (V, E)$ & a proximity graph with vertex set $V$ and edge set $E$ \\
$X_t$ & the dataset state at time $t$ \\
$X_t^G$ & the vector subset residing in GPU memory at time $t$ \\
$M$ & the GPU memory capacity (number of vectors) \\
$\lambda_x$ & the number of future accesses for vector $x$ \\
$F_{recent}(x, t)$ & the recent access count of vector $x$ at time $t$ \\
$E_{in}(x)$ & the number of in-neighbors for vector $x$ \\
$\alpha, \beta$ & the weight parameters in the prediction function \\
\hline
\end{tabular}
\vspace{-1.5em}
\end{table}

% The exact NNS requires computing the distances between a query and all vectors, which is computationally prohibitive for large datasets~\cite{wang21graphanns}. 
% To address this issue, most studies~\cite{faiss, NSG, hnsw, SSG} focus on approximately finding the nearest neighbors to achieve a balance between search accuracy and efficiency, called \emph{approximate nearest neighbor search} (ANNS).
Exact nearest neighbor search is computationally prohibitive for large datasets~\cite{wang21graphanns}, leading most studies~\cite{faiss, NSG, hnsw, SSG} to adopt \emph{approximate nearest neighbor search} (ANNS) for a balanced trade-off between accuracy and efficiency.
Given an approximation factor $\epsilon > 0$, an ANNS algorithm returns an ordered set of $k$ vectors $U_{\text{ANNS}}=\{x_1, x_2, \ldots, x_k\}$, sorted in ascending order of their distances to $q$. If $x_i^*$ is the $i$-th nearest neighbor of $q$ in $X$, the algorithm satisfies that $\text{dist}(x_i,q) \leq (1+\epsilon) \cdot \text{dist}(x_i^*,q)$ for all $i = 1, 2, \ldots, k$.
% The quality of the approximation is typically evaluated using the recall metric~\cite{PQ}, defined as 
% $\text{Recall@}k = \frac{|U_{\text{ANNS}} \cap U_{\text{NNS}}|}{|U_{\text{NNS}}|}$, 
% where $U_{\text{NNS}}$ denotes the ground truth set of the $k$ nearest neighbors. 
% A higher recall corresponds to a smaller $\epsilon$, thus, a higher degree of approximation, reflecting better search accuracy.

% \begin{equation}
% \label{eq:recall}
% \text{Recall@}k = \frac{|U_{\text{ANNS}} \cap U_{\text{NNS}}|}{|U_{\text{NNS}}|}
% \end{equation}

\subsection{Streaming ANNS}
\label{sec:streaming_anns}
The \emph{streaming approximate nearest neighbor search} (SANNS) problem extends traditional ANNS to handle continuously evolving datasets with dynamic insertions and deletions.
% Let $X_0 = \{x_1,\ldots, x_{N_0}\} \subset \mathbb{R}^D$ denote an initial vector dataset with $N_0$ vectors, where each vector has a unique identifier. 
SANNS processes an ordered sequence of operations $\mathcal{O} = \{o_1, o_2, \ldots, o_T\}$ over time, where $T$ is the number of time steps or operations in the stream. At each time step $t \in [1, T]$, exactly one operation $o_t$ is executed on the vector dataset state $X_{t-1}$, resulting in a new state $X_t$. 
It supports the following four types of operations: 

\begin{itemize}
    \item \textbf{Build}$(X_{\text{init}})$: $X_t = X_{\text{init}}$, initializing the dataset state, where $X_{\text{init}} \subset \mathbb{R}^D$, with periodic rebuilding optionally.
     % This operation constructs the initial index structure from a given static dataset, optionally used for periodic rebuilding.
     %Initializes the dataset $X_t = X_{\text{init}} \subset \mathbb{R}^D$ and constructs the initial index, optionally used for periodic rebuilding.
     
    \item \textbf{Search}$(q, k)$: $X_t = X_{t-1}$. Given a query vector $q$ and a positive integer $k$, this operation retrieves a set of $k$ vectors from $X_{t-1}$ that approximate the true $k$-nearest neighbors of $q$.
    
    \item \textbf{Insert}$(x_t)$: $X_t = X_{t-1} \cup \{x_t\}$, incorporating a new vector into the dataset. This operation dynamically expands the searchable vector space to accommodate new data.
    
    \item \textbf{Delete}$(x_t)$: $X_t = X_{t-1} \setminus \{x_t\}$, removing an existing vector $x_t \in X_{t-1}$. This operation contracts the search space by eliminating obsolete or irrelevant vectors.
\end{itemize}

In contrast to ANNS on static datasets with fixed ground truth, SANNS faces the challenges of preserving accuracy on evolving datasets while ensuring efficient dynamic operations.

\subsection{Graph-based index}
\label{sec:graph_index}
We define a directed graph $G = (V, E)$, where $V$ is the set of vertices with $|V| = N$, and each vertex $v_i \in V$ corresponds to a vector in the dataset. 
% The edge set $E \subseteq V \times V$ encodes pairwise connections between vertices whose associated vectors are close in the underlying vector space.
The edge set $E \subseteq V \times V$ encodes directional connections between vertices: $(v_i, v_j) \in E$ means vertex $v_i$ links to nearby vertex $v_j$, with edges directed from each vertex toward its selected neighbors.
For any vertex $v_i \in V$, we denote its out-neighbor set as $N_{out}(v_i) = \{v_j \in V \mid (v_i, v_j) \in E\}$ and its in-neighbor set as $N_{in}(v_i) = \{v_j \in V \mid (v_j, v_i) \in E\}$.

\textbf{Graph-Based ANNS and SANNS.}
Graph-based ANNS performs query processing by greedily traversing a proximity graph. 
Proximity graphs are primarily classified into two categories: \emph{k-nearest neighbor graphs (KNNG)} where each vertex maintains exactly $k$ outgoing edges to its approximate nearest neighbors~\cite{dong2011}, and \emph{relative neighborhood graphs (RNG)} that optimize connectivity by removing redundant edges while maintaining variable degrees~\cite{NSG,diskann}.
Given a query vector $q$, vector search begins from some entry points, which can be either randomly chosen~\cite{diskann,CAGRA,ParlayANN} or supplied by a separate algorithm~\cite{Iwasaki16,ADSampling}.
Then the procedure iteratively explores neighboring nodes with decreasing distance to $q$ until a stopping criterion is met~\cite{diskann}.
% This approach significantly reduces search complexity compared to exhaustive search while maintaining high recall.
To support streaming scenarios (SANNS), the index must handle continuous updates. At time $t$, let $G_t = (V_t, E_t)$ be the current graph. Given an insertion set $I_t$ and deletion set $D_t \subseteq V_t$, the graph is updated to $G_{t+1} = (V_{t+1}, E_{t+1})$, where
$V_{t+1} = (V_t \cup I_t) \setminus D_t$,
and
$
E_{t+1} \subseteq V_{t+1} \times V_{t+1}$.

\textbf{GPU Acceleration for Graph-based ANNS.} 
ANNS can be substantially accelerated using GPUs.
In practice, GPU-based methods typically adopt KNNG with fixed out-degree to fully exploit GPU's massive parallelism, as variable degrees would lead to load imbalance and underutilized computing resources~\cite{GGNN,CAGRA}. 
The key acceleration involves storing critical metadata such as graph structures in high-bandwidth memory, and utilizing one thread-block per query rather than one thread per query~\cite{GGNN} to enable extensive parallelization of core operations.
This design achieves substantial performance improvements over CPU-based approaches. 
% Details are provided in our extended version\footnote{\url{https://github.com/zjuDBSystems/svfusion}}.

\section{Overview}
\label{sec:overview}

We present SVFusion, a GPU–CPU–disk collaborative framework for SANNS. Our framework addresses the fundamental challenge of supporting real-time vector search and updates. 
To overcome GPU memory capacity limitations, (1) we introduce a \textbf{hierarchical graph-based vector index} that maintains synchronized graph structures across GPU, CPU, and disk tiers, enabling seamless transitions between memory layers as dataset size grows; (2) we design a \textbf{workload-aware vector placement strategy} that dynamically manages data residency and caching across GPU, CPU, and disk to minimize PCIe and I/O overheads. 
To further improve the performance under high-frequency updates, 
(3) we propose a \textbf{real-time coordination mechanism} with \textbf{concurrency control} that ensures consistency across memory tiers while enabling high-throughput, low-latency concurrent operations.
As shown in Figure~\ref{fig:system_architecture}a, SVFusion achieves these through three key components:

\begin{figure*}[t]
\centering
\includegraphics[width=0.92\linewidth]{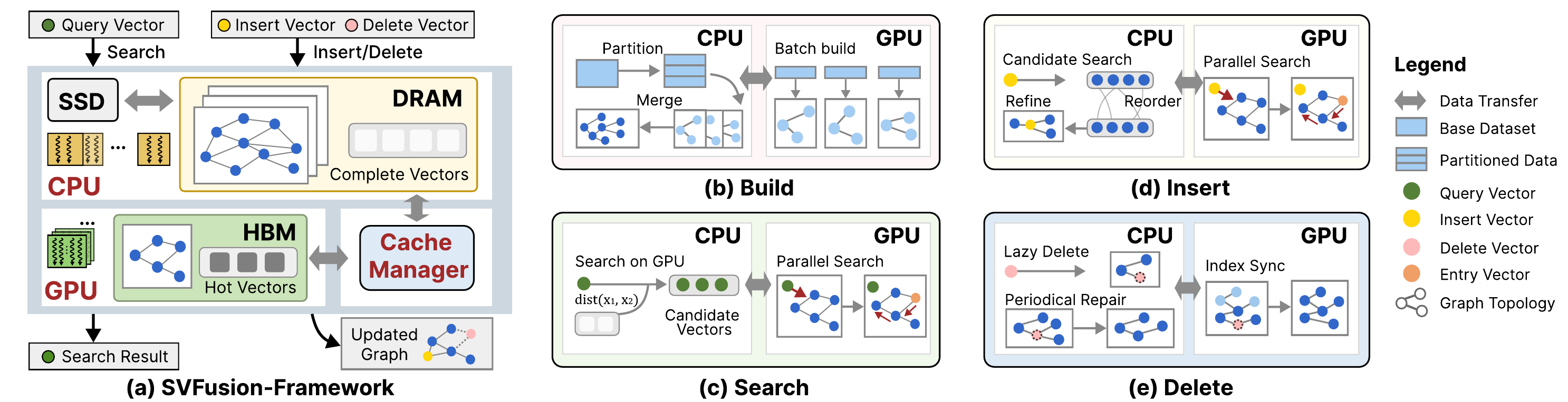}
\vspace{-1em}
\caption{Overview of SVFusion.}
\label{fig:system_architecture}
\vspace{-1em}
\end{figure*}

The \textbf{GPU module} accelerates vector search and updates by storing hot vectors and subgraphs in high-bandwidth memory (HBM). We extend CAGRA~\cite{CAGRA}, a state-of-the-art GPU-based index originally designed for static datasets,  into a dynamic, update-efficient index with specialized algorithms that support real-time vector search at scales exceeding GPU memory capacity.

The \textbf{CPU module} maintains the complete index graph with multi-version replicas and stores all vectors in DRAM or offloads cold vectors to disk when CPU memory is exhausted.
It orchestrates CPU computation and synchronization, and enables concurrent access control.

A dedicated \textbf{Cache Manager} manipulates data placement across GPU, CPU, and disk tiers to optimize query performance under hybrid workloads. Unlike traditional caching strategies that rely on temporal locality~\cite{LRU}, our cache manager exploits both vector access patterns and graph topology to dynamically load and evict vector entries, maximizing GPU memory residency for vectors accessed during graph traversal.

% These components enable flexible memory management across HBM, DRAM, and even disk storage, forming a scalable foundation for streaming vector operations. By decoupling storage and computation, the system supports adaptive data placement and efficient access across memory tiers. This hybrid design overcomes the memory limits of GPU-only systems and the performance bottlenecks of CPU-only systems, offering broad applicability to large-scale, real-time vector search scenarios.

\textbf{SANNS Workflow.}
\label{sec:system_workflow}
Figure~\ref{fig:system_architecture}b-\ref{fig:system_architecture}e illustrates SVFusion’s end-to-end SANNS workflow comprising the following four operations through coordinated CPU-GPU processing:

\underline{Build ($\S\ref{sec:hierarchical_structure}$)}.   
The index is constructed using a GPU-parallel strategy inspired by CAGRA~\cite{CAGRA}, performing neighbor search and graph linking to fully leverage GPU parallelism and memory bandwidth.
For large datasets, it partitions the data, builds subgraphs for each partition on GPU, and merges them into a unified index on CPU while preserving inter-partition connectivity.
When the complete graph exceeds CPU memory, this operation involves merging subgraphs within a bounded memory window that avoids loading the entire graph into memory.
% This design enables efficient GPU acceleration while scaling to datasets beyond memory constraints.

\underline{Search ($\S\ref{sec:hierarchical_structure}$ \& \ref{sec:avc})}. Given a query, this operation performs graph traversal starting on GPU. The \textbf{cache manager} checks whether required vectors or subgraphs are already cached in GPU memory.
If present, the search proceeds entirely on GPU; otherwise, SVFusion either transfers the data from CPU on demand or adaptively computes distances on CPU to reduce transfer overhead.
Finally, the top-$k$ results are obtained via sorting and returned to the user.

\underline{Insertion ($\S\ref{sec:insertion}$)}. This operation integrates new vectors while preserving graph structure and search quality.
Each vector first undergoes a GPU-based search to identify candidate neighbors, with on-demand transfers if needed. The CPU then performs \textbf{heuristic reordering} to refine neighbor selection and applies \textbf{reverse edge insertion} to maintain bidirectional connectivity.
% After a batch of insertions, updated subgraphs are synchronized from DRAM to HBM to ensure consistency, with large-batch transfers amortizing the overhead.
% To support fault tolerance and manage memory capacity, SVFusion also periodically persists the updated graph and vector data to disk.

\underline{Deletion ($\S\ref{sec:delete_handling}$)}. This operation adopts a lightweight deletion mechanism to handle large-scale vector deletion requests. It is based on two complementary strategies: (1) \textbf{lazy deletion}, which marks vectors as deleted without immediate structural changes; (2) \textbf{asynchronous repair}, which restores graph connectivity through localized topology-aware repair and periodic global consolidation. 
\section{Graph-based Vector Search with CPU-GPU Co-processing}
\label{sec:system_design}
To handle massive vector datasets with dynamic access patterns, SVFusion introduces a hierarchical index architecture, a workload-adaptive caching mechanism, and a real-time coordination strategy, enabling high-throughput, low-latency streaming vector search.

% Large-scale dynamic vector search scenarios require graph-based indices to handle massive datasets with continuously evolving index structures and access patterns.
% While CPU-GPU co-processing offers promising performance gains, effectively managing memory allocation, coordinating task execution, and handling dynamic data placement remains challenging.
% To this end, SVFusion introduces a hierarchical index architecture, a workload-adaptive caching mechanism, and a real-time coordination strategy to achieve high-throughput and low-latency streaming vector search.

\subsection{Motivation}
\label{sec:motivation}
\emph{Hierarchical Memory Management.} Due to the limited capacity of GPU memory, billion-scale vector datasets cannot be fully accommodated. 
While compressing vectors~\cite{FusionANNS,RUMMY} reduces memory footprint, it often degrades search accuracy. Observing that real-world access patterns are highly skewed~\cite{mohoney2025quake}, we adopt a hierarchical memory design, caching hot vectors uncompressed on GPU to enable scalable, high-accuracy search across massive datasets.
% \textbf{Motivation.}
% In large-scale and dynamic vector search scenarios, systems must handle massive datasets with continuously evolving index structures and access patterns due to real-time insertions and deletions. To maintain high throughput and high accuracy under dynamic workloads, several key factors must be considered in CPU-GPU co-processing design:

\emph{Dynamic Cache Management:} In streaming ANNS, access patterns evolve continuously, making previously hot vectors cold.
Unlike traditional caching approaches~\cite{LRU,lrfu} that fetch data before computation, vector access in CPU-GPU co-processing determines both where computation occurs and whether data transfer is needed.
This motivates a workload-aware cache mechanism to optimally adapt vector placement under evolving workloads.

% \emph{Hybrid Processing Coordination:} 
% CPU-GPU collaborative ANNS involves multiple computational stages~\cite{FusionANNS} across heterogeneous processors.
% However, effective coordination remains challenging due to synchronization overhead and resource underutilization.
% This calls for a pipeline design that minimizes synchronization while enabling concurrent execution.

% \changeone{
% \emph{Real-time Coordination for Heterogeneous Workloads.} As illustrated in Table 1, streaming ANNS workloads exhibit two distinct requirements.
% Although existing GPU-based ANNS methods leverage large-batch processing to fully utilize GPU resources, achieving high throughput, they cause unacceptable delays for individual queries, and simply reducing to small-batch processing underutilize the GPU resources.
% This motivates a real-time coordination mechanism that dynamically adapts execution strategies to achieve both high throughput and low latency.
% }

\subsection{Hierarchical Graph-based Vector Index}
\label{sec:hierarchical_structure}
\begin{figure}[t]
\centering
\includegraphics[width=0.92\linewidth]{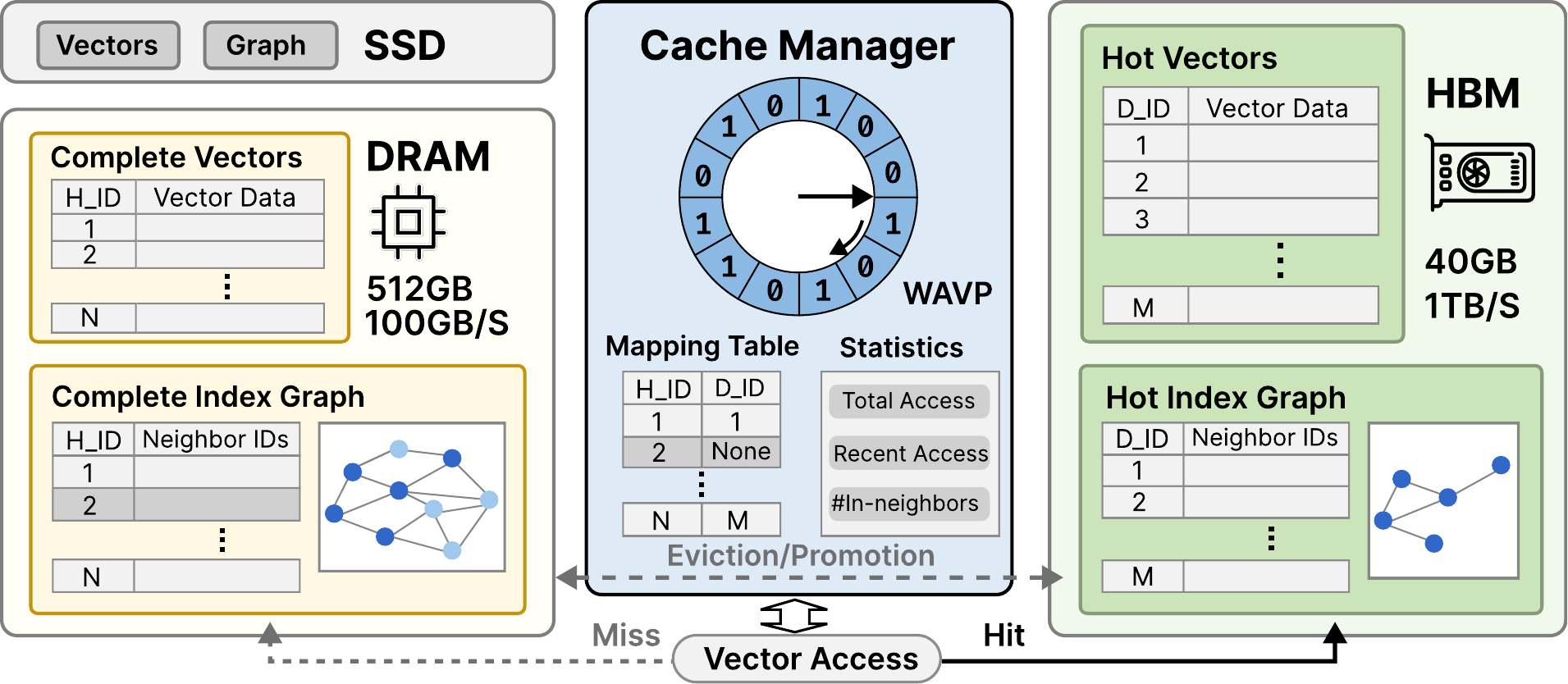}
\vspace{-1em}
\caption{Hierarchical Index Structure in SVFusion.}
\label{fig:two_tiered_index}
\vspace{-1em}
\end{figure}

Figure~\ref{fig:two_tiered_index} illustrates our hierarchical index structure that spans CPU and GPU memory tiers. Unlike existing GPU-accelerated ANNS methods that are constrained by limited GPU memory capacity when handling large-scale datasets~\cite{CAGRA,SONG}, our design leverages the complementary characteristics of CPU and GPU memory: GPU memory provides high bandwidth (1 TB/s) but limited capacity, while CPU memory offers large capacity (hundreds of GB) but lower bandwidth (100 GB/s).
To bridge this bandwidth-capacity gap, our hierarchical design organizes vector data across multiple storage tiers, with a co-processing search algorithm that dynamically leverages both processors to achieve high-throughput ANNS on large-scale datasets.

\textbf{Hierarchy Index Structure.}
Our index adopts a three-tier GPU–CPU–disk architecture, enabling efficient caching, seamless data movement, and scalability beyond GPU and main memory limits.
Given a dataset of size $N$, the CPU memory maintains the graph structure $G = (V, E)$ with host-side identifiers ($h\_id \in [1, N]$).
The GPU memory caches a selected subset of $M \ll N$ hot vectors with compact device-side identifiers ($d\_id \in [1, M]$). 

By maintaining hot vectors and their associated subgraphs on GPU, most computations can be served directly from high-bandwidth memory without accuracy loss.
For large-scale datasets exceeding DRAM capacity, our framework seamlessly extends to incorporate persistent storage, creating a three-tier hierarchy. 
Vectors and graph structures are stored on disk in the same format as in CPU memory, with frequently accessed portions cached in CPU and GPU tiers through on-demand loading.
This design enables scaling from millions to billions of vectors while maintaining consistent performance for varying workloads.

The \textbf{cache manager} resides on GPU to coordinate data movement across these storage tiers during SANNS operations. 
It maintains a mapping table with the mapping function $mapping(h\_id) \rightarrow \{d\_id \mid \text{NONE}\}$, which returns the device identifier for GPU-cached vectors or \textit{NONE} for CPU-resident vectors.
This design leverages high-bandwidth GPU memory access for massive parallel vector lookups without the overhead of CPU-GPU synchronization.

\textbf{Index Construction.}  
During index construction, the dataset is partitioned based on GPU memory capacity. When extended to persistent storage with large-scale datasets, SVFusion loads data of each partition from disk to CPU, transfers it to GPU for local subgraph construction, and writes the completed subgraph back to disk. The CPU then merges these subgraphs by computing cross-partition edges within a bounded memory window, loading vectors on demand from disk. This approach ensures memory usage regardless of dataset scale, with the final graph persisted to disk while frequently accessed portions are cached in CPU-GPU tiers.

\begin{algorithm}[t]
\caption{ANNS using CPU-GPU Co-Processing}
\label{alg:svfusion}
\begin{algorithmic}[1]
\Require{graph index $G$, a batch of $p$ queries $Q = \{q_1, q_2, \ldots, q_p\}$, $k$ for top-$k$, and candidate pool size $L \geq k$}
\Ensure{$K := \cup_{i=1}^{p}\{K_i\}$, where $K_i$ is the set of $k$NNs for $q_i$}
\For{each query $q_i \in Q$ \textbf{in parallel}}
\State $\mathcal{C}_i \gets$ InitCandidatePool($G$, $L$) \Comment{random initialization}
% \While{not converged}
\Repeat
    \State $x_{curr} \gets \mathcal{C}_i.\text{GetNearest}(q_i)$
    \State $X_i \gets$ FetchNeighbors($x_{curr}$, $G$) \Comment{on GPU}
    \State $X_i^{GPU}, X_i^{CPU} \gets \emptyset, \emptyset$ \Comment{initialize processing set}
    \For{each $x \in X_i$ \textbf{in parallel}}  
        \State $d\_id \gets \text{mapping}(x)$ \textbf{or} WAVP($x$)
        \\ \Comment{vector placement called only on cache miss}
        \State $X_i^{GPU} \gets X_i^{GPU} \cup \{x | d\_id \neq \text{NONE}\}$
        \State $X_i^{CPU} \gets X_i^{CPU} \cup \{x | d\_id = \text{NONE}\}$
    \EndFor
    \State $D_i^{CPU} \gets$ ParallelComputeDist($q_i$, $X_i^{CPU}$) 
    \Comment{on CPU}
    \State $D_i^{GPU} \gets$ ParallelComputeDist($q_i$, $X_i^{GPU}$) 
    \Comment{on GPU}
    \State $\mathcal{C}_i.\text{Update}(D_i^{CPU} \cup D_i^{GPU},X_i)$  
    \Comment{update candidates}
\Until{$\mathcal{C}_i$ unchanged from previous iteration}
\State $K_i \gets$ $k$ nearest candidates to $q_i$ from $\mathcal{C}_i$
\EndFor
\end{algorithmic}
\end{algorithm}

\textbf{CPU-GPU Vector Search.}
We present SVFusion with the search workflow in Algorithm~\ref{alg:svfusion}.
Unlike prior GPU-only approaches, SVFusion prioritizes GPU execution for cached data and dynamically offloads uncached computations to the CPU or triggers on-demand transfers. 
The parallel search process begins with randomly initializing each query’s candidate pool (Lines 1-2).
Then the graph traversal is performed on GPU (Line 5) until the candidate pool remains unchanged.
Random entry points provide GPU-friendly parallelism and avoid bias seed maintenance under dynamic updates.
During neighbor expansion (Lines 7-11), our cache manager checks each neighbor vector's cache status (Line 8) and partitions computation on CPU or GPU. 
On cache hits, vectors are processed directly on GPU using high-bandwidth memory access. For cache misses, we propose a Workload-Aware Vector Placement (WAVP) algorithm that determines whether to promote vectors to GPU or perform computations directly on CPU. 
This results in partitioning neighbors into $X_i^{GPU}$ and $X_i^{CPU}$ for parallel distance computation on both processors (Lines 10-13). 
This adaptive strategy balances throughput and memory efficiency, enabling scalable and responsive vector search under GPU memory constraints. 
Note that maintaining data copies across CPU and GPU introduces consistency challenges during index updates. We address this through immediate bitset synchronization for logical deletions and batched graph updates from CPU to GPU, detailed in Section~\ref{sec:cc_update}.
 
\subsection{Workload-Aware Vector Placement}
\label{sec:avc}
To address the dynamic access patterns of streaming workloads, we propose an adaptive vector placement strategy for optimizing data locality and efficient computation during ANNS.
%Rather than static allocation, it prioritizes vectors based on run-time access statistics to maintain high query throughput under varying workload conditions.

Let $X_t = \{x_1, x_2, \ldots, x_N\}$ denote the full set of vectors at timestamp $t$. Due to GPU memory constraints, only a subset $X_t^G \subseteq X_t$ with $|X_t^G| = M \ll N$ resides in GPU memory, while the remaining vectors are kept in CPU memory.
When accessing a vector $x \notin X_t^G$, the \textbf{core challenge} of CPU-GPU co-processing is to decide the optimal placement for each vector access: (1) promoting vectors from CPU to GPU memory, or (2) executing distance computations directly on CPU. 
This fundamentally differs from traditional cache replacement policies, such as LRFU~\cite{lrfu}, which are optimized to maximize cache hit ratios.
In CPU–GPU co-processing, promoting every missed vector to GPU cache might lead to excessive data transfers, while computing all queries on CPU underutilizes GPU parallelism.
Our caching strategy instead jointly optimizes transfer cost and computational efficiency, deciding adaptively whether to cache a vector on GPU or to compute directly on CPU.

To determine the optimal choice between these two strategies for each vector access not resident on GPU,  we need to quantify the trade-off between computational savings and data transfer overhead. We define the following gain function:
\begin{equation}
\label{eq:score}
\text{gain}(x) = \lambda_x \cdot (T_{\text{CPU}} - T_{\text{GPU}}) - T_{\text{transfer}}
\end{equation}

Here, $\lambda_x$ represents the number of future accesses for vector $x$ within a specified time window. $T_{\text{CPU}}$ and $T_{\text{GPU}}$ represent the average time required to compute distances on CPU and GPU, respectively. 
$T_{\text{transfer}}$ is the amortized cost of transferring a vector to GPU memory, typically calculated based on a batch size of 2048 to reduce the communication overhead per vector~\cite{cuda_guide}.
When $\text{gain}(x) > 0$, it implies that the expected cumulative speedup from GPU acceleration outweighs the cost of data transfer, justifying the promotion of $x$ to GPU memory.

The key challenge in implementing this gain function lies in obtaining $\lambda_x$ during ANNS. 
Since future access frequency is inherently unobservable at decision time, we develop a prediction function $F_{\lambda}$ to estimate $\lambda_x$ based on observable runtime information.

\begin{figure}[t]
    \centering   
    \begin{subfigure}{0.495\linewidth}
        \centering
        \includegraphics[width=\linewidth]{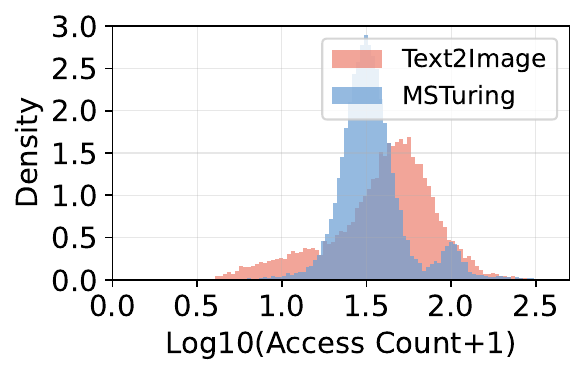}\
        \vspace{-2em}
        \caption{Distribution of Recent Access.}
        \label{fig:score_analysis_1}
    \end{subfigure}
    \begin{subfigure}{0.49\linewidth}
        \centering
        \includegraphics[width=\linewidth]{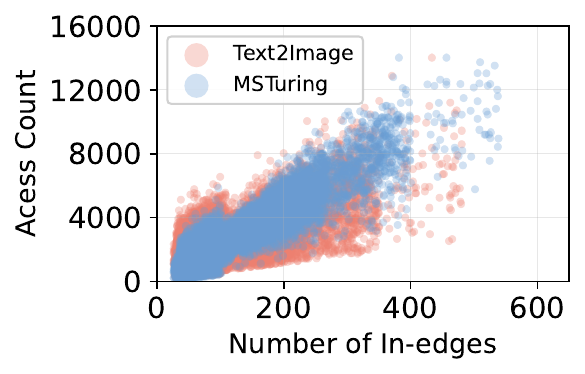}\
        \vspace{-2em}
        \caption{Distribution of In-neighbors.}
        \label{fig:score_analysis_2}
    \end{subfigure}
    \vspace{-2.5em}
    \caption{Characterization of Vector Access Patterns.}
    \label{fig:score_analysis}
\vspace{-1em}
\end{figure}

\textbf{Prediction Function Design.}
An effective prediction of $\lambda_x$ requires properties that correlate with future access. As graph-based ANNS relies on graph traversal, access patterns naturally depend on both temporal factors and graph structure. 
We therefore analyzed vector access with respect to both properties during ANNS across MSTuring~\cite{msturing} and Text2Image~\cite{text2image} with different dimensions and data types.
Figure~\ref{fig:score_analysis} illustrates the key characteristics of vector access patterns. Recent access frequency $F_{recent}(x, t)$ captures temporal locality but exhibits a medium-hot distribution rather than a long-tail. In contrast, in-neighbor degree $E_{in}(x)$ shows a strong structural correlation with access frequency, reflecting that searches often traverse high-connectivity hubs.
However, such structural importance cannot be exploited by existing cache strategies such as LRFU~\cite{lrfu} that bases its replacement decision on recency and frequency. 
To effectively cache these high-connectivity vectors in the graph, our method explicitly incorporates graph structural connectivity into the caching decision process.
% This suggests we should prioritize caching these high-connectivity vectors in GPU memory to maximize cache hit rates.

Based on these observations, we design a prediction function to estimate future access counts by combining temporal locality $F_{recent}(x, t)$ and structural connectivity $E_{in}(x)$ as key factors.
We adopt a linear combination approach for its simplicity and interpretability, though other fitting methods (e.g., polynomial) could also be explored:
\begin{equation}
\label{eq:predict}
F_\lambda(x) = \alpha \times 
F_{recent}(x, t)
+
\beta \times \log(1 + E_{in}(x))
\end{equation}
where parameters $\alpha$ and $\beta$ control the relative importance of temporal and structural features, respectively. In practice, we fix the ratio between $\alpha$ and $\beta$ and tune their values based on performance evaluation.
The cache manager maintains $F_{recent}(x, t)$ using a sliding window mechanism that periodically decays older accesses, and updates $E_{in}(x)$ as the graph topology evolves.

\textbf{Theoretical Analysis.} 
We provide theoretical justification that our prediction function can effectively guide caching decisions. 
We model the actual future access count $\lambda_x$ as:
% We model the future access count $\lambda_x$ as a linear combination of observable features:
\begin{equation}
\lambda_x = \lambda_1 \cdot F_{recent}(x,t) + \lambda_2 \cdot \log(1 + E_{in}(x)) + \lambda_3
\end{equation}
where $\lambda_1, \lambda_2 > 0$ are weight parameters and $\lambda_3$ is a bias term. The logarithmic transformation is used to diminish extremely high values of $E_{in}(x)$.
By setting $\alpha = \lambda_1$ and $\beta = \lambda_2$, we have $\lambda_x = F_{\lambda}(x) + \lambda_3$. 
% This indicates that ranking vectors by $F_{\lambda}(x)$ is equivalent to ranking by expected future accesses.
Recall that caching is beneficial when $\text{gain}(x) > 0$. It holds when $\lambda_x > \rho$, where $\rho=\frac{T_{\text{transfer}}}{T_{\text{CPU}}-T_{\text{GPU}}}$ is the cost ratio of transfer overhead to computation gain. Since we have $\lambda_x = F_{\lambda}(x) + \lambda_3$, the condition becomes $F_{\lambda}(x) + \lambda_3 > \rho$, which is equivalent to $F_{\lambda}(x) > \rho - \lambda_3$. By setting $\tau = \rho - \lambda_3$, we obtain $F_{\lambda}(x) > \tau \Longleftrightarrow \lambda_x > \rho \Longleftrightarrow \text{gain}(x) > 0$. This shows our approach reduces cache decisions to simple threshold comparisons.

\textbf{The WAVP Algorithm}. Based on the prediction function, we design the workload-aware vector placement algorithm illustrated in Algorithm~\ref{alg:avc}.
Given a vector $x$ not currently in GPU memory, the algorithm returns either the device identifier for caching or \emph{NONE} to indicate CPU-based execution. The placement algorithm consists of two phases: \emph{selective prefetching} and \emph{predictive replacement}.

\begin{algorithm}[!t]
\caption{Workload-Aware Vector Placement for Cache Miss}
\label{alg:avc}
\begin{algorithmic}[1]
\Require vector $x$ not resident in GPU, vector subset in GPU Memory $X_t^G$, clock pointer $clock$, reference bits $ref$
\Ensure a valid device identifier $d\_{id}$ or $\text{NONE}$
\If{$F_{\lambda}(x) \leq \theta$}
  \State \Return $\text{NONE}$
\EndIf
\State $F_{min} \leftarrow \min\{F_{\lambda}(x_i) \mid x_i \in X_t^G, ref[i] = 0\}$
\While{true}
  \State $x_{curr} \leftarrow X_t^G[clock]$
  \If{$ref[clock] = 0$ \textbf{and} $F_{\lambda}(x_{curr}) = F_{min}$}
      \State $\text{replace}(x_{curr}, x)$
      \State \Return $\text{mapping}(x_{curr})$
  \ElsIf{$ref[clock] = 1$}
      \State $ref[clock] \leftarrow 0$
  \EndIf
  \State $clock \leftarrow (clock + 1) \bmod |X_t^G|$
\EndWhile
\end{algorithmic}
\end{algorithm}

The \textit{selective prefetch} phase (Lines 1-2) determines whether to promote vector $x$ to GPU memory. We initialize the threshold $\theta = \frac{T_{\text{transfer}}}{T_{\text{CPU}} - T_{\text{GPU}}}$, which corresponds to the minimum frequency required for GPU promotion to be beneficial. When $F_{\lambda}(x) \leq \theta$, the system keeps $x$ on CPU for in-place computation. 
In the \textit{predictive replacement} phase (Lines 3-11), we extend the clock-sweep policy with a prediction-guided eviction strategy: when GPU memory is full, the system evicts the vector with zero reference bits and the lowest predicted frequency $F_{\lambda}(x)$, effectively mitigating thrashing under dynamic ANNS workloads.
This design strictly controls eviction decisions to avoid unnecessary vector replacements and redundant data transfers.

% The \textit{predictive replacement} phase (Lines 3-11) extends a clock-sweep mechanism with prediction-guided victim selection when GPU memory is full.
% As discussed above, access patterns in ANNS workloads make traditional LRU-based methods prone to thrashing, where specific vectors are evicted but re-fetched in a short interval.
% We address this by incorporating predicted access frequency into victim selection: among vectors with zero reference bits, we evict the one with the minimum $F_{\lambda}(x)$ (Line 6). 

To scale to large-scale datasets, SVFusion extends its memory hierarchy to include disk storage, forming a cascading lookup pipeline across GPU, CPU, and disk. Upon cache misses, vectors are asynchronously prefetched from lower tiers, guided by a hash-based directory that tracks data residency. When CPU memory is saturated, vectors with the lowest $F_\lambda$ scores are demoted to disk, ensuring efficient multi-tier data flow and consistent performance at scale.

\subsection{Real-time Coordination for SANNS}
\label{sec:coordination}
To support both high-throughput and low-latency workloads, we design a real-time coordination mechanism that dynamically adapts GPU execution. 
While large batches improve throughput, latency-sensitive queries are processed immediately via a low-latency execution path combining multi-stream coordination~\cite{RTAMS-GANNS} and multi-CTA execution~\cite{CAGRA}. 
An adaptive resource manager further adjusts GPU utilization under fluctuating streaming loads, and cold-start handling ensures stable performance from system initialization.

\textbf{Low-latency Processing.}
To achieve low latency for small-batch queries in latency-oriented scenarios, we leverage two techniques to utilize GPU resources.
First, we integrate a CUDA multi-stream coordination mechanism, comprising multiple search streams and a dedicated update stream.
This design enables concurrent operations: search streams process queries of varying batch sizes in parallel, while the update stream handles insertions and deletions asynchronously with a consistency guarantee detailed in \S\ref{sec:cc_update}.
Second, inspired by CAGRA's multi-CTA design~\cite{CAGRA}, the system enables intra-query parallelism where multiple GPU thread blocks collaborate for a single query.

\textbf{Adaptive Resource Management.}
SVFusion dynamically allocates GPU resources to adapt to varying workloads.
To support concurrent multi-stream execution, we partition GPU memory into independent segments guarded by lightweight spinlocks, with vectors mapped to segments via identifier hashing.
Streams with higher workload overhead are automatically allocated more GPU memory and compute resources.
Additionally, the caching threshold $\theta$ is continuously adjusted based on observed miss rates and the distribution of estimated future access counts $F_\lambda(x)$. When miss rates increase along with higher average $F_\lambda$, $\theta$ is increased to prevent cache thrashing by being more selective about replacements.
All adaptations operate automatically without manual tuning.

\textbf{Cold Start Handling.} 
To avoid startup slowdowns when the system begins processing streaming operations, SVFusion performs a three-phase initialization: (1) \textbf{GPU resource priming} pre-allocates vector storage in GPU global memory and creates CUDA streams for asynchronous execution; (2) \textbf{data structure initialization} prepares the cache manager's mapping table and begins to access tracking metadata; and 
(3) \textbf{GPU cache warm-up} preloads vectors based on predicted access counts. 
For large datasets, SVFusion caches top-ranked high in-degree vectors predicted by Equation~\ref{eq:predict} to minimize GPU cache misses.

\section{Index Update}
\label{sec:index_update}
\subsection{Vector Insertion}
\label{sec:insertion}
The insertion process consists of two phases: (1) it performs a search operation to find candidate neighbors for new vectors, and (2) then establishes edges from the new vector to its selected neighbors and updates these neighbors with reverse edges.
To achieve high throughput, our framework leverages GPU's massive parallelism to enable large batches of insertions concurrently, compared to CPU-based methods that are limited to around hundreds of vectors due to thread pool constraints~\cite{upreti2025cost}.
For low-latency requirements, we employ the techniques described in \S\ref{sec:coordination}, such as multi-stream and multi-CTA, to maintain high GPU utilization.

We further design a rank-based candidate reordering strategy. 
For each candidate at position $i$ in the candidate list with sequential order, we count how many previously selected neighbors (at position $j < i$) contain this candidate in their neighbor lists—each occurrence represents a detourable path that can bypass the direct edge. We then sort candidates by detourable path counts in ascending order and select the top candidates up to the graph degree. This avoids expensive distance computations with only $\mathcal{O}(|\mathcal{C}| \log |\mathcal{C}|)$ complexity to traverse the candidate set $\mathcal{C}$.
During reverse edge insertion, we employ fine-grained atomic operations and thread-local buffers to handle concurrent graph updates, achieving better scalability than coarse-grained locking while preserving consistency.
%The details are presented in the Appendix.

\subsection{Vector Deletion}
\label{sec:delete_handling}
SVFusion handles vector deletions through a three-stage process: (1) logical marking of inactive vertices to avoid query disruption, (2) localized topology-aware repair for rapid connectivity recovery, and (3) periodic global consolidation asynchronously.

\subsubsection{Logical Deletion.}
SVFusion employs a lightweight, lazy deletion strategy to efficiently support high-frequency removal operations.
Instead of immediately modifying the graph structure, SVFusion marks deleted vertices as logically removed using a bitset.
These marked vertices are transparently skipped during subsequent insertions, deletions, and queries.
By deferring the costly operations of structural repair and neighbor updates, this strategy minimizes deletion latency and avoids synchronization bottlenecks.

\subsubsection{Affected Vertex Repair.}
\begin{figure}[!t]
    \centering   
    \begin{subfigure}{0.43\linewidth}
        \centering
        \includegraphics[width=\linewidth]{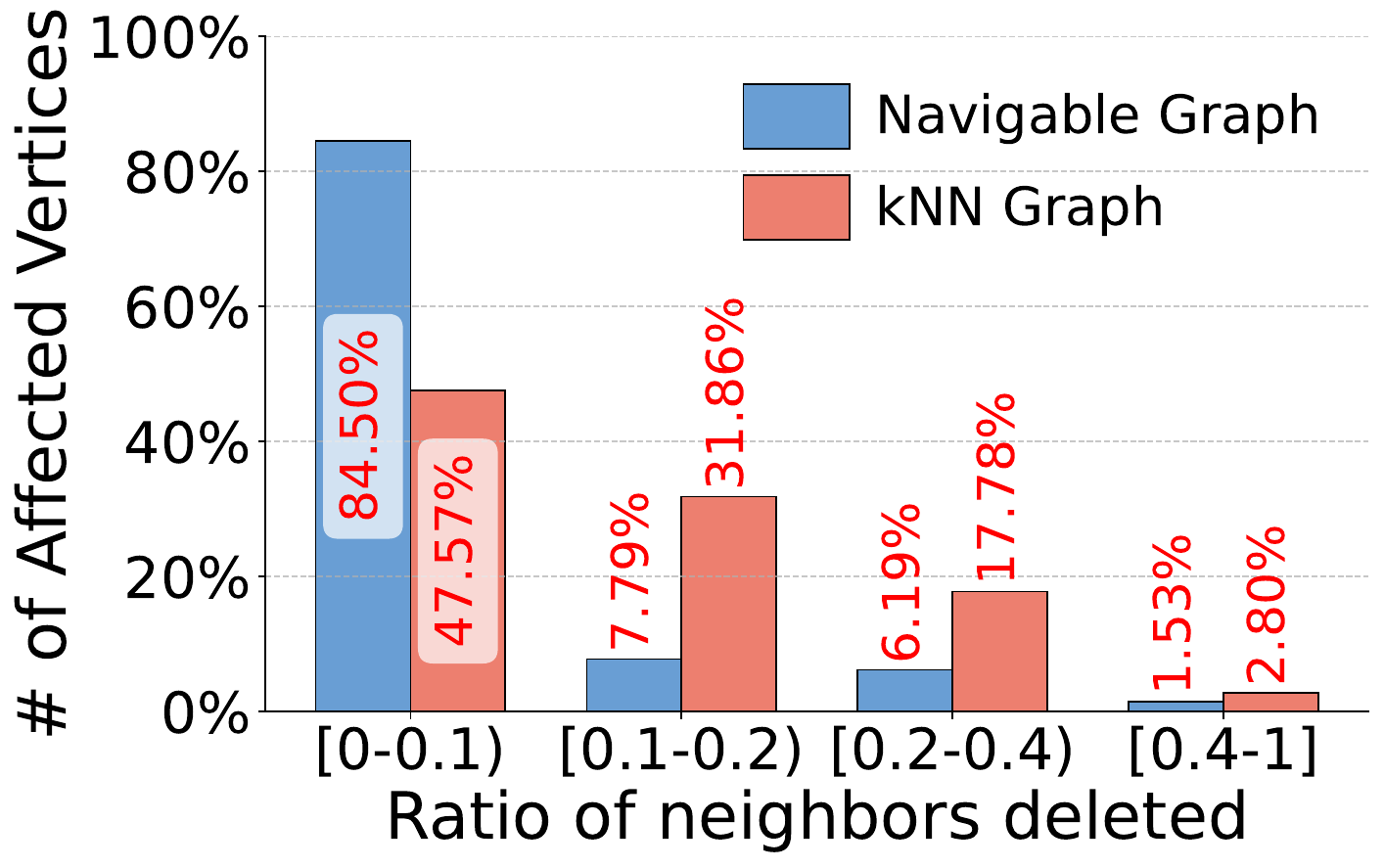}\
        \vspace{-2em}
        \caption{Wikipedia}
        \label{fig:exp_cost2}
    \end{subfigure}
    \begin{subfigure}{0.43\linewidth}
        \centering
        \includegraphics[width=\linewidth]{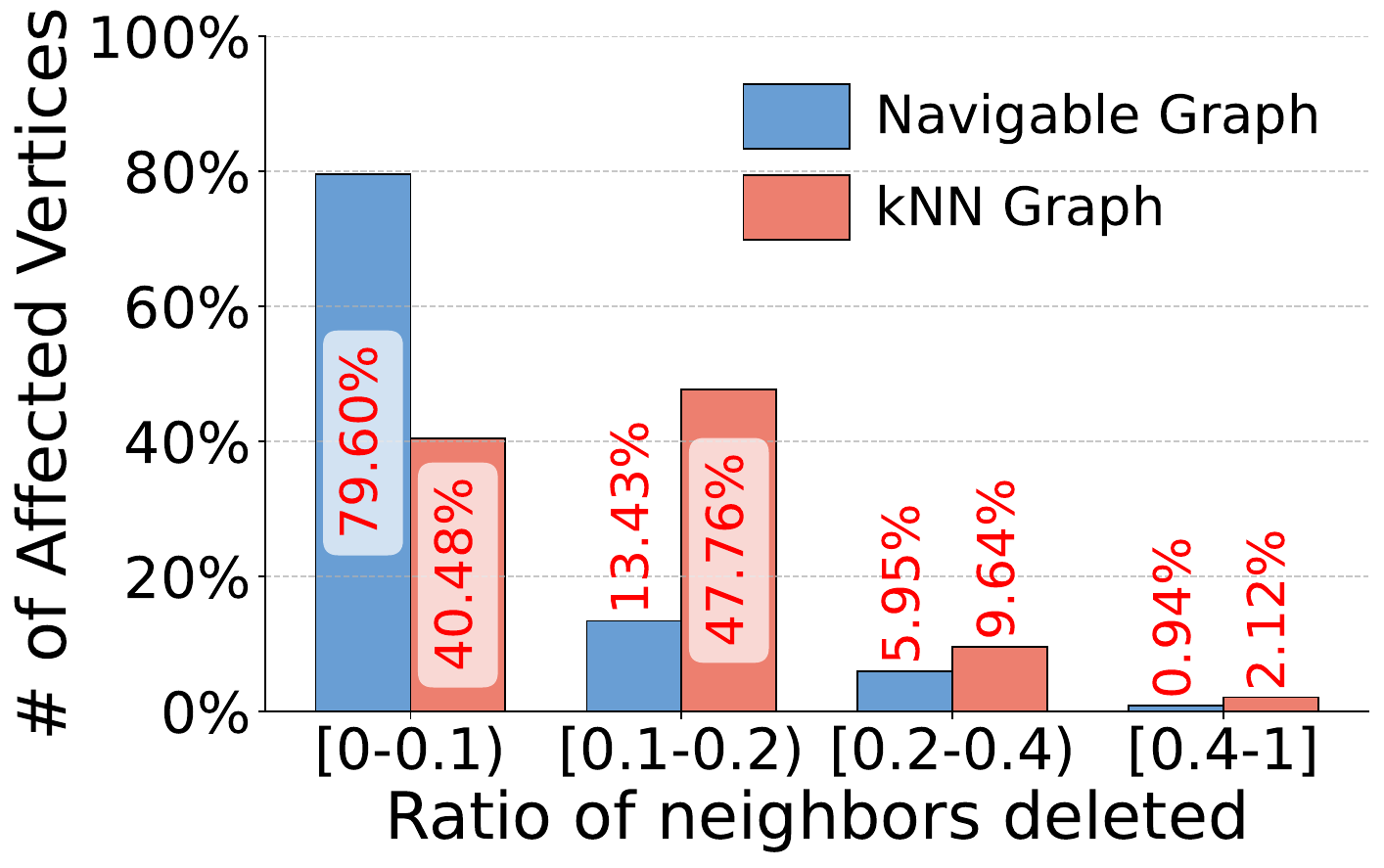}\
        \vspace{-2em}
        \caption{MSMARCO}
        \label{fig:exp_cost3}
    \end{subfigure}
    \vspace{-1em}
    \caption{Distribution of Deleted Neighbor Counts.}
    \label{fig:delete_distribution}
\vspace{-1.5em}
\end{figure}

While logical deletions enable low-latency removals, their accumulation gradually fragments the graph structure, degrading connectivity and search efficiency.
Previous deletion handling methods~\cite{freshdiskann} are primarily designed for RNG~\cite{NSG}, where deletions primarily affect high-degree hubs and global consolidation is deferred until a deletion threshold (e.g., 20\%) is reached.
However, KNNG enforces uniform degree constraints, causing deletion impacts to spread evenly across all vertices. 
Our analysis (Figure~\ref{fig:delete_distribution}) shows that after deleting 10\% of vertices, KNNG has 3.26$\times$ more vertices with 10-40\% deleted neighbors and 2.04$\times$ more with over 40\% deleted neighbors.
This pervasive connectivity loss degrades search quality long before global repair thresholds are reached, necessitating continuous, incremental repairs focused on the most affected vertices as deletions accumulate.

% To quantify this structural difference, we conduct an experimental analysis comparing deletion impact distributions. As shown in Figure~\ref{fig:delete_distribution}, after deleting 10\% of vertices, $k$NN graphs exhibit significantly higher neighbor deletion ratios: the proportion of affected vertices with 10-40\% deleted neighbors is 3.26$\times$ higher, and those with over 40\% deleted neighbors are 2.04$\times$ higher compared to navigable graphs. 
% Therefore, widespread connectivity degradation severely affects search quality long before global repair thresholds are reached, necessitating continuous incremental repair that targets critical vertices as deletions accumulate.

\textbf{Localized Topology-Aware Repair.}
SVFusion addresses the problem of rapid degradation through a localized topology-aware repair strategy. 
Rather than immediately cleaning up deleted vertices, the approach leverages the deletion bitset to quickly identify severely affected vertices $V^L$ where more than 50\% of neighbors have been deleted, and applies lightweight repair only to these critical vertices.
% expanding the candidate set with a subset of the deleted vertices' neighbors, followed by robust pruning~\cite{NSG} with relaxed filtering conditions. This incremental approach maintains graph quality with minimal overhead, as it processes only a small number of severely affected vertices rather than all vertices impacted by deletion.
Given a vertex $v \in V^L$, in contrast to the expensive consolidation that connects all vertices in $N_{out}(p)$ to $v$ for each deleted neighbor $p$, our lightweight repair selects at most $c$ vertices from $N_{out}(p)$ to connect to $v$. The constant $c$ is small (we use $c=8$), ensuring the number of added edges is $O(cR)$, which is much smaller than the $O(R^2)$ during consolidation, where $R$ is the graph degree. 
% \textcolor{red}{After adding the selected edges in distance order, we apply robust pruning\cite{NSG} with relaxed filtering conditions to maintain the neighbor count of $v$ constrained by $R$.
% }
This incremental approach preserves graph quality with minimal overhead compared to performing expensive repairs on all affected vertices.

\textbf{Global Consolidation.}
% When the deletion ratio surpasses a predefined threshold (e.g., 20\%), SVFusion globally consolidates all affected neighborhoods by aggregating candidates from outgoing neighbors of the deleted vertices.
% This consolidation runs asynchronously in the background with the latest snapshot and new updates, allowing concurrent insertions and searches to continue without interruption. To synchronize with foreground operations, the process operates on a graph snapshot, with a graph merging mechanism handling the coordination between different graph versions.
When the deletion ratio surpasses a predefined threshold (e.g., 20\%), SVFusion performs a global consolidation of all affected neighborhoods by aggregating candidates from the outgoing neighbors of deleted vertices. This consolidation executes asynchronously in the background on the latest graph snapshot, allowing insertions and queries to proceed concurrently without disruption. To ensure consistency with sequential operations, the process leverages a graph snapshot and a merging mechanism to coordinate updates across different graph versions.

\subsection{Concurrency Control}
\label{sec:cc_update}
SVFusion employs a two-level concurrency control protocol to enable safe and efficient SANNS operations across hierarchical memory tiers. 
We first establish \emph{consistency guarantees} through fine-grained locking and synchronized data structures, ensuring searches observe consistent data despite interleaved updates. 
Building on this foundation, a \emph{multi-version mechanism} 
decouples expensive background consolidation from serial operations via snapshot isolation, enabling concurrent execution without blocking.

\textbf{Consistency Guarantees.} 
SVFusion provides a consistent view of the index across GPU, CPU, and disk tiers, ensuring that all queries and updates observe a coherent state of the hierarchical index. 
Consistency is maintained through fine-grained local synchronization and coordinated cross-tier updates over three shared structures: the CPU-resident graph, the multi-tier deletion bitset, and the CPU–GPU mapping table.
% This is achieved through two mechanisms: local synchronization within each processor and coordination across CPU-GPU memory. 
% The data structures that operations synchronize over are the graph structures on CPU, the deletion bitset maintained on all memory tiers, and the CPU-GPU mapping table. 
\begin{figure}[!t]
\centering
\includegraphics[width=0.8\linewidth]{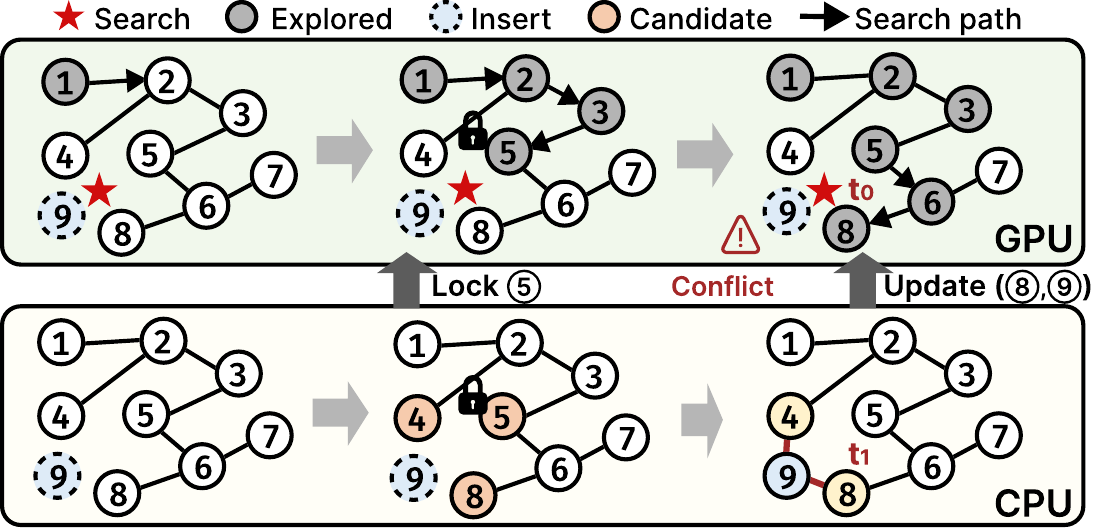}
\vspace{-1em}
\caption{Illustration of the Concurrency Control Workflow.}
\label{fig:cc_control}
\vspace{-1.5em}
\end{figure}

\underline{Local synchronization.} 
Within each memory tier, fine-grained locks coordinate concurrent operations. 
Searches traverse neighbor lists under read locks, while insertions acquire read locks for candidate search and upgrade to write locks only on affected neighbor lists during edge establishment. 
Deletions update the shared bitset under exclusive lock while marking the affected version, allowing other operations to skip logically deleted vertices.

% Within each memory space, we employ fine-grained locking to coordinate concurrent operations. Searches traverse neighbor lists using read locks.
% % while insertions acquire write locks on affected subgraphs for the entire batch duration. 
% Insertions acquire read locks during candidate search, then write locks only on the neighbor lists being modified during edge establishment.
% Deletions update the bitset under exclusive lock, leaving other operations to independently check the bitset to skip deleted vectors.
% % Background consolidation operates on snapshots, acquiring write locks only during the brief merge phase.

\underline{Cross-tier coordination.}
To maintain consistency during updates, SVFusion adopts a two-phase synchronization protocol between CPU and GPU memory.
For deletions, bitsets are atomically updated across tiers under exclusive lock to ensure consistent visibility.
For topology updates (insertions, repairs, or consolidations), modified neighbor lists are first committed on the CPU under write lock with an incremented version, while corresponding GPU cache entries retain their versions.
Subsequently, these updated lists are asynchronously propagated to the GPU in batches with version identifiers updated once the transfer completes.
During this transient phase, GPU queries encountering version conflicts transparently fall back to CPU memory to ensure consistent results.
Based on this lightweight version control mechanism, asynchronous updates never overwrite concurrently modified regions, preserving correctness and avoiding synchronization stalls.
This design guarantees correctness under concurrent operations while enabling high-throughput, non-blocking propagation across hierarchical memory tiers.
Figure~\ref{fig:cc_control} illustrates concurrent insertion and search in SVFusion, ensuring correct visibility of new vectors via fine-grained locking and cross-tier synchronization.

% Maintaining data consistency across CPU-GPU memory during updates is necessary to prevent searches from observing inconsistent states.
% For vector deletions, we atomically update bitsets under exclusive lock at both sides.
% For topology changes (insertion, localized repair, or consolidation), we apply a two-phase strategy: modified neighbor lists are first committed to the CPU graph under write lock with corresponding GPU cache entries immediately marked as invalid to prevent reading outdated data on GPU. 
% The updated neighbor lists are then asynchronously transferred to GPU in batches and marked valid upon completion. This design ensures consistency while avoiding blocking on cross-memory transfers.

\textbf{Multi-version Mechanism.} 
To decouple expensive background consolidation from online vector operations, SVFusion adopts a multi-version concurrency mechanism that maintains a small set of consistent graph versions across memory tiers. Each version provides a complete, immutable snapshot of the graph topology for concurrent background processing, while new operations proceed on the active version under consistency guarantees.

A new graph version is created whenever a background task (e.g., consolidation or large-scale repair) begins. 
The system duplicates the current graph metadata as a read-only snapshot $G_{t_0}$ for task execution. Insertions and deletions continue on the active graph $G_{t_1}$. Upon completion, the background task produces an updated graph $G'_{t_0}$, which is merged back into $G_{t_1}$ to form a unified version.
To avoid unbounded version accumulation, SVFusion enforces a bounded-version policy, deferring new version creation once the limit is reached and scheduling consolidation serially.

% As described in $\S\ref{sec:delete_handling}$, SVFusion employs a multi-version design where background graph consolidation operates on snapshots, while \changeone{searches and insertions continue on the active graph subject to our consistency guarantee}.
% Our multi-version mechanism works by creating a new version whenever a background task (e.g., consolidation) starts, duplicating the current graph as a read-only snapshot for task processing.
% Upon task completion, the results are merged back into the active graph. This background merging creates additional versions for concurrent updates, potentially triggering further merge operations.
% To prevent unbounded memory growth, we limit concurrent versions and defer new updates when this threshold is reached.
% Without proper synchronization, however, this merging process can lead to data inconsistency. We therefore propose a lightweight two-phase merging algorithm while maintaining high update throughput.

\underline{Incremental Subgraph Appending.} 
During merging, we first identify new vertices inserted after the snapshot was taken:
$V_{new} = V(G_{t_1}) \setminus V(G_{t_0})$.
The corresponding induced subgraph $G_{new} = G_{t_1}[V_{new}]$ is directly appended to the consolidated result $G'_{t_0}$.
This incremental approach ensures that recently inserted vertices are retained in the merged structure without reprocessing the entire graph, substantially reducing merge latency.

% When consolidation begins at time $t_0$ using snapshot $G_{t_0}$, ongoing insertions update the active graph to $G_{t_1}$ by $t_1$. 
% We identify newly inserted vertices $V_{new} = V(G_{t_1}) \setminus V(G_{t_0})$ and directly append the corresponding subgraph $G_{new} = G_{t_1}[V_{new}]$ to the consolidated graph $G'_{t_0}$.

\underline{Reverse Edge Integration.} 
To maintain bidirectional connectivity, we record reverse edge updates that occur between existing and newly inserted vertices during the consolidation window $[t_0, t_1]$. Each update is logged as a triplet $(v, v_{new}, d)$,  where $v\in V(G_{t_0})$, $v_{new}\in V_{new}$, and $d$ denotes their precomputed distance.

Before committing the new version, SVFusion atomically applies these updates by augmenting each affected vertex neighborhood with the corresponding reverse edges.
This two-phase strategy ensures both structural completeness and connectivity preservation in the merged index.

\section{Evaluation}
\label{sec:evaluation}

% In this section, we first introduce the evaluation setup and then present a detailed analysis of the evaluation results.

\subsection{Experimental Setup}
\label{sec:exp_setup}

\textbf{Hardware Configuration.}
The experiments are conducted on a dual-socket server running Ubuntu 18.04.6 LTS, equipped with two Intel Xeon Gold 5218 CPUs (32 physical cores, 64 threads, forming two NUMA nodes), 376 GB DDR4 DRAM (12 channels, 256 GB/s), an NVIDIA A100 GPU with 40 GB device memory (PCIe 3.0), and Intel D3-S4510 960 GB enterprise SSDs for persistent storage.

\begin{table}
\caption{Statistics of experimental datasets.}
\label{tab:datasets}
\vspace{-1.5em}
\footnotesize
\centering
\begin{tabular}{|c|c|c|c|c|c|}
\hline
\textbf{Datasets} & \textbf{N} & \textbf{Size (GB)} & \textbf{D} & \textbf{$\mathbf{N_q}$} & \textbf{Type} \\ 
\hline
Wikipedia~\cite{wikipedia} & 35,000,000 & 101 & 768 & 5,000 & Text \\
\hline
MSMARCO~\cite{msmarco} & 101,070,374 & 293 & 768 & 9,376 & Text\\
\hline
MSTuring~\cite{msturing} & 200,000,000 & 75 & 100 & 10,000 & Text\\
\hline
Deep1B~\cite{deep1b} & 1,000,000,000 & 358 & 96 & 10,000 & Image\\
\hline
Text2Image~\cite{text2image} & 100,000,000 & 75 & 200 & 10,000 & Image\\
\hline
\end{tabular}
\vspace{-1.5em}
\end{table}

\noindent
\textbf{Datasets.} 
We evaluate SVFusion on five widely used real-world datasets with diverse dimensions and data types, which have been extensively utilized~\cite{annbenchmarks,simhadri2024results,mohoney2025quake}.
Table~\ref{tab:datasets} summarizes the detailed properties of each dataset.
Queries are randomly sampled from each dataset, with ground truth computed via exhaustive linear scan.

\noindent
\textbf{Workloads.} 
Our evaluation covers diverse streaming workload patterns, each consisting of a collection of vectors and an operation sequence
as defined in \S\ref{sec:streaming_anns}. The workloads differ in: (\romannumeral 1) the interval between insertions and deletions, and (\romannumeral 2) the spatial correlation among vectors deleted in a single step. 
To simulate realistic streaming scenarios, we continuously consume search, insertion, and deletion vectors from a Kafka source at varying QPS rates.
We adopt three representative workload patterns from the \emph{2023 Big ANN Challenge}~\cite{simhadri2024results} and \emph{MSTuring-IH} generated in ~\cite{mohoney2025quake}:
\begin{itemize}[leftmargin=0pt, itemindent=10pt]
    \item \textbf{SlidingWindow}: The dataset is evenly divided into $T_{max} = 200$ segments. One segment is inserted per step, and from step $T = T_{max}/2 + 1$, the segment inserted $T_{max}/2$ steps earlier is deleted. Evaluation begins at $T_{max}/2 + 1$. This workload simulates a sliding window that retains only the most recent data.
    
    \item \textbf{ExpirationTime}: Vectors are assigned different lifetimes—short-term (10 steps), long-term (50 steps), and permanent (100 steps)—in a 10:2:1 ratio. At each step, $1/T_{max}$ of the dataset is inserted, with expired vectors removed accordingly. This setup evaluates index behavior under heterogeneous data lifespans.
    
    \item \textbf{Clustered}: The dataset is partitioned into 64 clusters using k-means across 5 rounds. Each round alternates between insertion and deletion phases: randomly inserting and then deleting vectors from each cluster. As spatially adjacent points are modified together, this workload creates strong distributional shifts.

    \item \textbf{MSTuring-IH}: Starting with 20 million vectors, the dataset grows to 200 million over 1,000 operations with a 90\% insert and 10\% search ratio, testing scalability and query stability under large-scale incremental growth.
\end{itemize}

% Beginning with 20 million vectors, the dataset grows to 200 million as we process 1,000 operations with a 90\% insert and 10\% search ratio. This tests the ability to handle large-scale growth while maintaining query quality.
% 

% \textbf{Implementation Details.}
\noindent
\textbf{Implementation and Baselines.}
We implement SVFusion by extending the cuVS 24.02 library~\cite{cuvs} following the architecture in Figure~\ref{fig:system_architecture}. 
% Implementation details are available in our GitHub repository\footnote{\url{https://github.com/zjuDBSystems/svfusion}}.
We compare SVFusion with the following baselines:
\begin{itemize}[leftmargin=0pt, itemindent=10pt]
    \item \textbf{HNSW}~\cite{hnsw} is a representative graph-based ANNS method, extensively deployed in industrial systems and studied in academia. We use $M=48$, and $ef\_construction=ef\_search=128$.
    \item \textbf{FreshDiskANN}~\cite{freshdiskann} is the state-of-the-art graph-based ANNS method that supports dynamic scenarios on memory or SSDs. 
    % It adopts an out-of-place update strategy by incrementally repairing the graph structure in the index. 
    We configure FreshDiskANN with the settings in~\cite{nips_2023_streaming}: $R=64, l_b=l_s =128, \alpha = 1.2$ for insertion, and $l_d = 128$ for search.

    \item \textbf{CAGRA}~\cite{CAGRA} is NVIDIA's high-performance ANNS algorithm designed for GPUs. It achieves superior performance but encounters GPU memory limitations for large-scale datasets.

    \item \textbf{PilotANN}~\cite{PilotANN} is a recent hybrid architecture that effectively enables GPU acceleration, but does not support dynamic updates.

    \item \textbf{GGNN}~\cite{GGNN} is a GPU-based ANNS method that constructs hierarchical graph structures similar to HNSW, but is designed for static datasets that fit within GPU memory capacity.
\end{itemize}
% For our large-scale streaming vector search scenario, existing GPU-based methods typically support only static construction~\cite{SONG,RUMMY,CAGRA} or provide limited insertion capabilities constrained by GPU memory~\cite{RTAMS-GANNS}, with no open-source availability. To fairly evaluate the CPU-GPU co-processing capabilities of SVFusion, we extend both FreshDiskANN and HNSW with GPU-accelerated versions that offload distance computations to the GPU.

Existing GPU-based methods either support only static construction~\cite{SONG,RUMMY,CAGRA} or have limited dynamic capabilities constrained by GPU memory~\cite{RTAMS-GANNS}. For fair comparison on streaming workloads, we create GPU-accelerated versions of FreshDiskANN and HNSW by offloading distance computations to GPU, and extend GGNN to support updates with GPU cache management.

\noindent
\textbf{Evaluation Metrics.} 
% The design goal of SVFusion is to achieve efficient vector updates while ensuring high search accuracy.
We focus on four key metrics: 
(1) \textbf{Throughput}. It measures the number of search queries and update operations (insertions/deletions) processed per second under streaming workloads; (2) \textbf{Latency}. It denotes the end-to-end time from when a request is submitted until the final result is returned; (3) \textbf{Recall}. Recall@$k$ denotes the recall for each search query, with $k$=$10$ by default; 
(4) \textbf{GPU cache miss rate.} It is defined as the percentage of vector accesses during ANNS that require fetching data from CPU memory rather than the GPU cache. A lower miss rate indicates better cache residency for frequently accessed vectors.

\subsection{SANNS Overall Performance}
\label{sec:sanns_performance}

\begin{figure*}[t] 
\centering
  \includegraphics[width=0.92\linewidth]{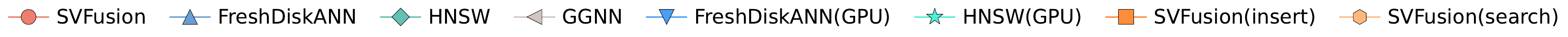}
  
  \begin{minipage}[t]{\textwidth}    
    \begin{minipage}[c]{0.035\textwidth}
      \raggedleft
      \rotatebox{90}{\small\parbox{2.5cm}{\centering Wikipedia-\\SlidingWindow}}
    \end{minipage}%
    \hspace{0.4em} 
    \begin{minipage}[c]{0.95\textwidth}
      \includegraphics[width=\linewidth]{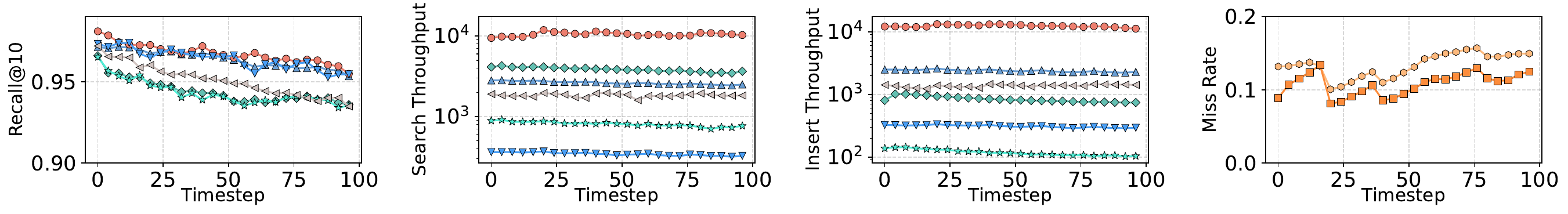}
    \end{minipage}
  \end{minipage}
  \vspace{-1em}
  
  \begin{minipage}[t]{\textwidth}      
    \begin{minipage}[c]{0.035\textwidth}
      \raggedleft
      \rotatebox{90}{\small\parbox{2.5cm}{\centering MSMARCO-\\Expiration}}
    \end{minipage}%
    \hspace{0.6em}%  
    \begin{minipage}[c]{0.95\textwidth}
      \includegraphics[width=\linewidth]{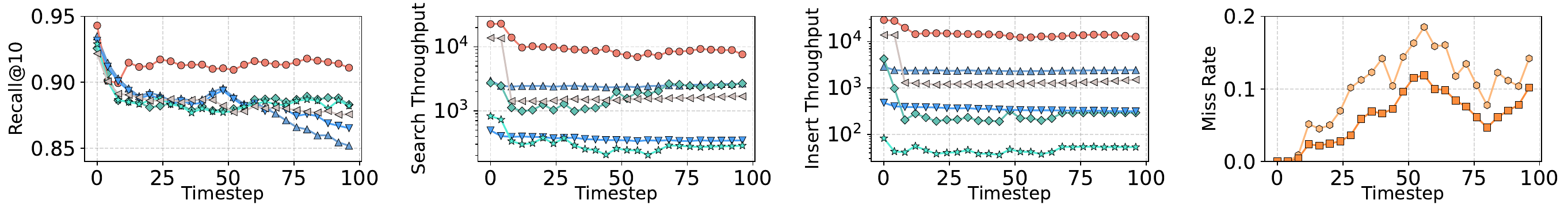}
    \end{minipage}
  \end{minipage}
  \vspace{-1em}

  \begin{minipage}[t]{\textwidth}      
    \begin{minipage}[c]{0.035\textwidth}
      \raggedleft
      \rotatebox{90}{\small\parbox{2.5cm}{\centering MSTuring-\\Clustered}}
    \end{minipage}%
    \hspace{0.6em}%  
    \begin{minipage}[c]{0.95\textwidth}
      \includegraphics[width=\linewidth]{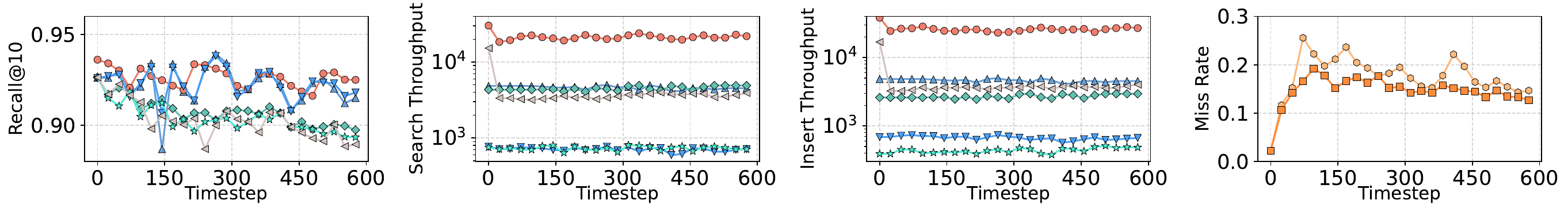}
    \end{minipage}
  \end{minipage}
  \vspace{-1em}

  \begin{minipage}[t]{\textwidth}      
    \begin{minipage}[c]{0.035\textwidth}
      \raggedleft
      \rotatebox{90}{\small\parbox{2.5cm}{\centering MSTuring-\\IH}}
    \end{minipage}%
    \hspace{0.6em}%  
    \begin{minipage}[c]{0.95\textwidth}
      \includegraphics[width=\linewidth]{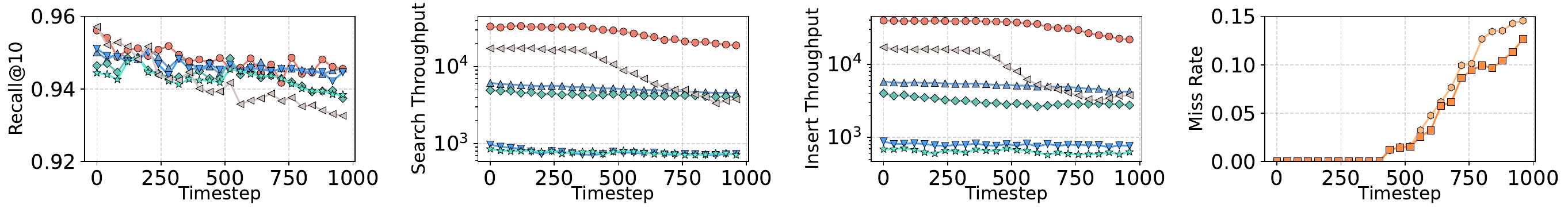}
    \end{minipage}
  \end{minipage}
  \vspace{-1em}

  \begin{minipage}[t]{\textwidth}      
    \begin{minipage}[c]{0.035\textwidth}
      \raggedleft
      \rotatebox{90}{\small\parbox{2.5cm}{\centering Text2Image-\\Clustered}}
    \end{minipage}%
    \hspace{0.6em}%  
    \begin{minipage}[c]{0.95\textwidth}
      \includegraphics[width=\linewidth]{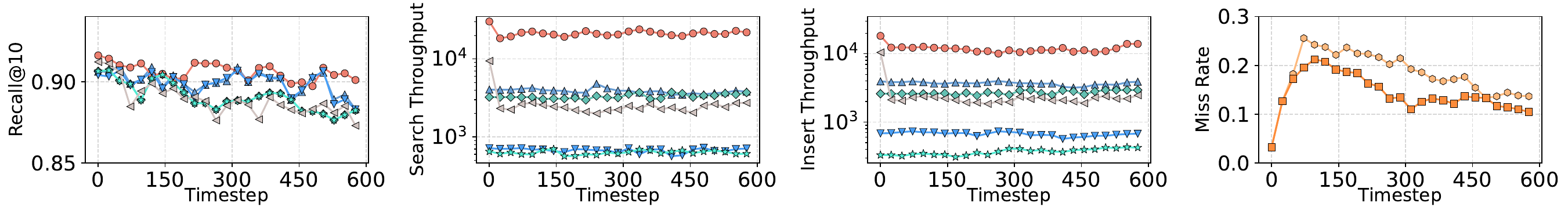}
    \end{minipage}
  \end{minipage}
  \vspace{-2em}
  
  \caption{Comparison of recall, search throughput, insert throughput, and miss rate for five various workloads.}
  \label{fig:sanns_performance}
\vspace{-1em}
\end{figure*}

To evaluate the overall performance of SVFusion, Figure~\ref{fig:sanns_performance} presents recall, search throughput, insert throughput, and GPU miss rate across seven workloads on five datasets.
These workloads involve high update volumes, with performance targets of high throughput and bounded p99 latency within 900~ms~\cite{Vearch}.
Delete throughput is not included since all methods use logical markers for deferred deletions, making instantaneous rates incomparable.
% Therefore, Table~\ref{tab:performance} provides a comprehensive breakdown of total execution time across all operations.
% Overall, SVFusion achieves superior performance across all operations:  up to 9.5$\times$ higher search throughput, 71.8$\times$ higher insertion throughput, and 9.59$\times$ faster deletion than FreshDiskANN, while maintaining the highest recall rates.
Overall, SVFusion consistently outperforms all baselines, achieving up to 9.5$\times$ higher search throughput and 71.8$\times$ higher insertion throughput than FreshDiskANN, while maintaining the highest recall throughout.

\textbf{Recall Analysis.} SVFusion maintains consistently high search quality across all streaming workloads, achieving Recall@10 scores between 91\% and 96\%, comparable to or surpassing all baselines. 
It outperforms FreshDiskANN with an average recall improvement of 0.4–3.4\%, attributed to its \textbf{localized topology-aware repair mechanism} that proactively detects and restores critical graph connectivity, mitigating structural degradation over time. In contrast, HNSW yields the lowest recall due to its lack of a deletion repair strategy, leading to persistent graph fragmentation.
All methods exhibit recall fluctuations under continuous structural updates, particularly in clustered workloads.
% where concentrated deletions amplify local connectivity disruptions.
% This improvement stems from our \textbf{localized topology-aware repair mechanism}, which continuously detects and fixes critical connectivity issues in the graph topology immediately, preventing the accumulation of structural degradation.
% In contrast, HNSW exhibits the lowest recall among all methods due to the absence of any repair mechanism during deletions, resulting in persistent structural fragmentation.
% We also observe that all methods exhibit recall fluctuations due to continuous graph structure updates. These fluctuations are particularly pronounced in the clustered workload, where concentrated regional deletions can severely impact local connectivity. 

\textbf{Throughput Analysis.} Figure~\ref{fig:sanns_performance} demonstrates that SVFusion achieves the highest throughput for both search and insertion operations, delivering on average 20.9$\times$ higher search throughput and 3.5$\times$ higher insertion throughput than FreshDiskANN and HNSW.
Notably, GPU-accelerated baselines, such as GGNN run 4.5–7.8$\times$ slower than SVFusion once the dataset exceeds GPU capacity, as frequent CPU–GPU transfers negate the benefits of GPU acceleration.
SVFusion sustains high performance through its adaptive caching algorithm.
During early timesteps of workloads initialized from empty datasets (\emph{ExpirationTime}, \emph{Clustered} and \emph{MSTuring-IH}), search throughput peaks at 40K due to fully utilized GPU computation and minimal cache misses (<1\%).

\textbf{Miss Rate Analysis.} 
We also report GPU cache miss rates for insertion and search operations to better characterize multi-tier memory behavior. 
Despite GPU memory representing only a small fraction of the dataset, SVFusion maintains consistently low miss rates across dynamic workloads, demonstrating the effectiveness of its cache placement strategy in capturing temporal and graph structural access patterns and minimizing redundant evictions.

\begin{figure}[!t] 
\centering
  \includegraphics[width=\linewidth]{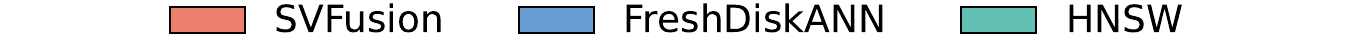}
 \begin{minipage}[t]{\linewidth}    
    \begin{minipage}[c]{0.015\linewidth}
      \raggedleft
      \rotatebox{90}{\footnotesize\parbox{2cm}{\centering \textbf{QPS=500}}}
    \end{minipage}%
    \hspace{0.1em} 
    \begin{minipage}[c]{0.985\linewidth}
      \includegraphics[width=\linewidth]{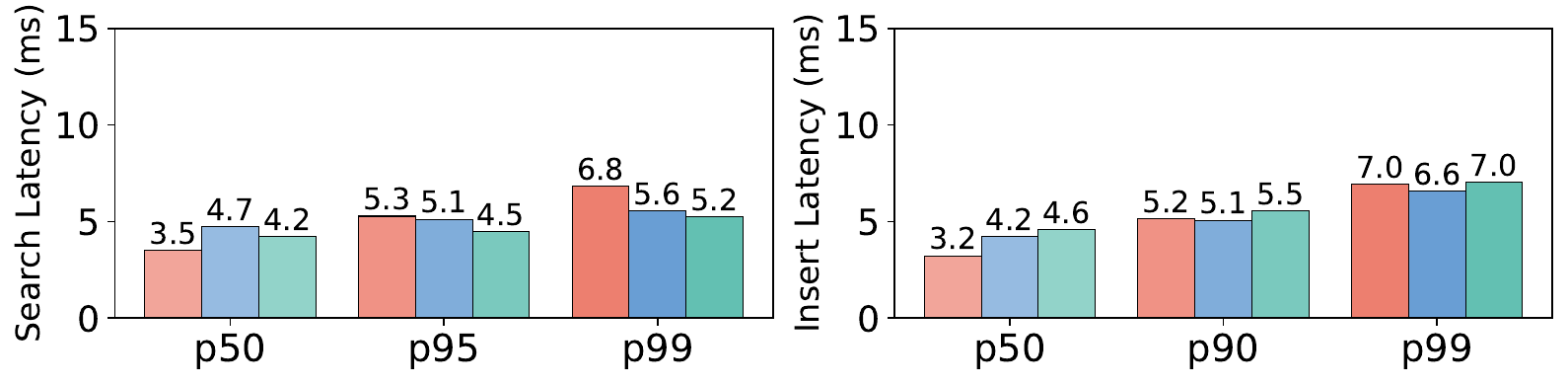}
    \end{minipage}
  \end{minipage}
  \vspace{-0.5em}

  \begin{minipage}[t]{\linewidth}    
    \begin{minipage}[c]{0.015\linewidth}
      \raggedleft
      \rotatebox{90}{\footnotesize\parbox{2cm}{\centering \textbf{QPS=2000}}}
    \end{minipage}%
    \hspace{0.1em} 
    \begin{minipage}[c]{0.985\linewidth}
      \includegraphics[width=\linewidth]{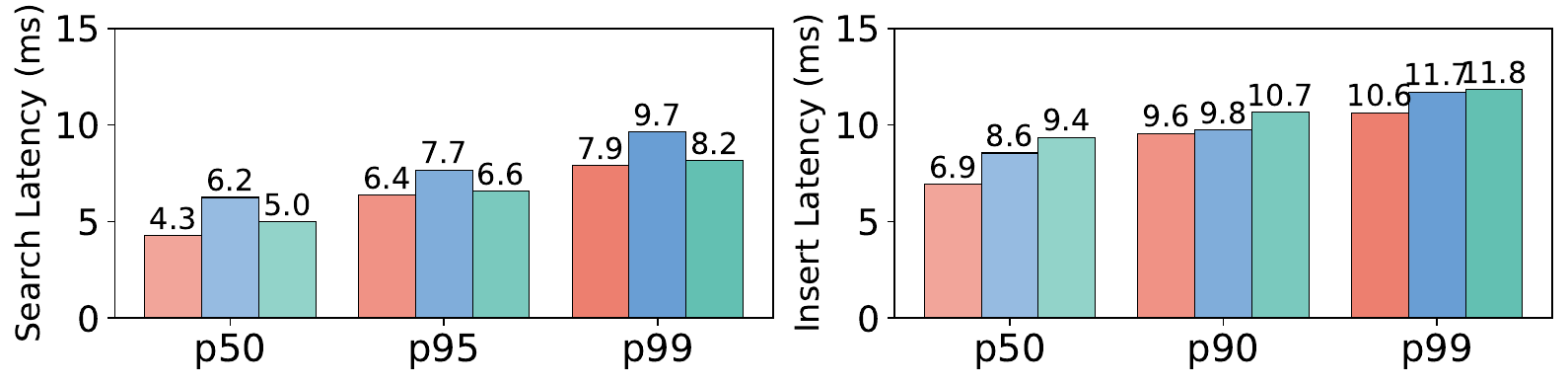}
    \end{minipage}
  \end{minipage}
  \vspace{-0.5em}

  \begin{minipage}[t]{\linewidth}    
    \begin{minipage}[c]{0.015\linewidth}
      \raggedleft
      \rotatebox{90}{\footnotesize\parbox{2cm}{\centering \textbf{QPS=10000}}}
    \end{minipage}%
    \hspace{0.1em} 
    \begin{minipage}[c]{0.985\linewidth}
      \includegraphics[width=\linewidth]{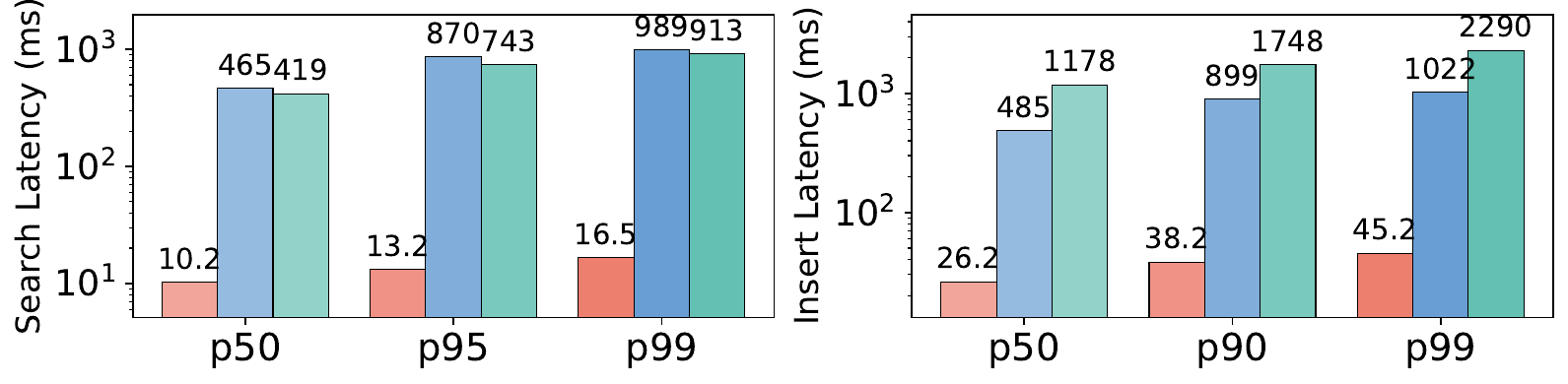}
    \end{minipage}
  \end{minipage}

  \vspace{-1em}
\caption{Comparison of p50, p95 and p99 query latencies.}
\vspace{-1.5em}
\label{fig:sanns_lantency}
\end{figure}

\textbf{Latency Analysis.}
To comprehensively evaluate SVFusion’s latency performance under varying streaming intensities, we scale the request rate from low (QPS=500) to high (QPS=10,000) on the MSTuring dataset.
Figure~\ref{fig:sanns_lantency} reports the p50, p95, and p99 latencies for both search and insertion operations.
At low request rates, SVFusion delivers latency on par with baseline systems, with a moderate 15–30\% increase in p99 latency due to CPU–GPU synchronization overhead that cannot be effectively amortized under limited query concurrency. Despite this effect, all observed latencies remain below 10 ms, indicating minimal real impact on responsiveness.
Under moderate throughput, SVFusion demonstrates strong latency stability, achieving p50 search latency of 4.3 ms and p99 of 7.9 ms, representing 32\% and 18\% reductions compared to FreshDiskANN and HNSW, respectively.
This improvement results from SVFusion’s multi-stream coordination mechanism, which overlaps computation and data transfer to mitigate blocking across concurrent operations.
At high request rates, the advantage becomes even more pronounced: while baseline methods experience severe queue buildup with p99 latencies exceeding 900 ms (45–50.7$\times$ slower), SVFusion sustains tail latencies of 16.5 ms for search and 45.3 ms for insertion.
These results highlight SVFusion’s scalable coordination design, achieving both high throughput and consistently low latency, whereas CPU-bound baselines saturate rapidly under heavy load due to limited parallelism and resource contention.

\subsection{Component Analysis}
\label{sec:exp_components}
% We conduct a systematic evaluation of SVFusion’s key components to quantify their respective contributions to overall performance.

\begin{figure*}[t] 
\centering
  \includegraphics[width=0.92\textwidth]{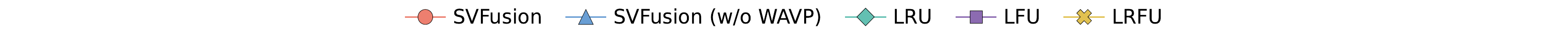}
  \includegraphics[width=\textwidth]{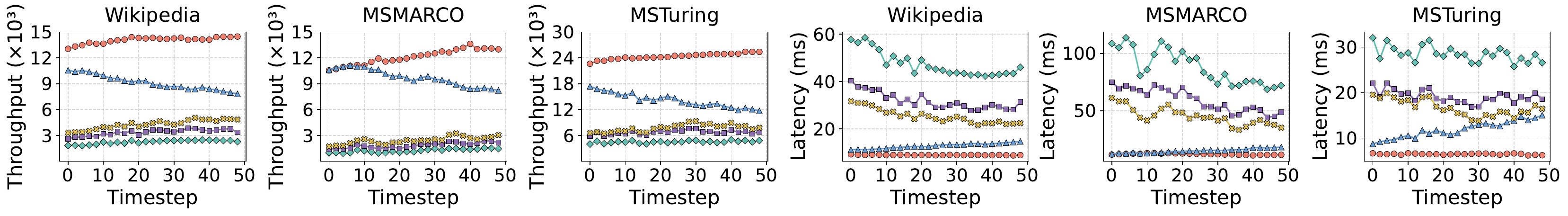}
  \vspace{-2em}
  \caption{Comparison of different replacement strategies on update throughput and latency.}
  \label{fig:replacement_comparison}
\vspace{-1em}
\end{figure*}

\begin{figure*}[t] 
\centering
  \includegraphics[width=0.92\textwidth]{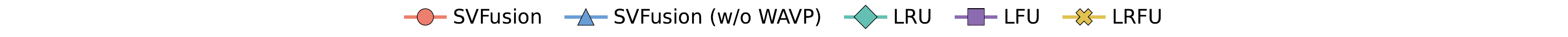}
  \includegraphics[width=\textwidth]{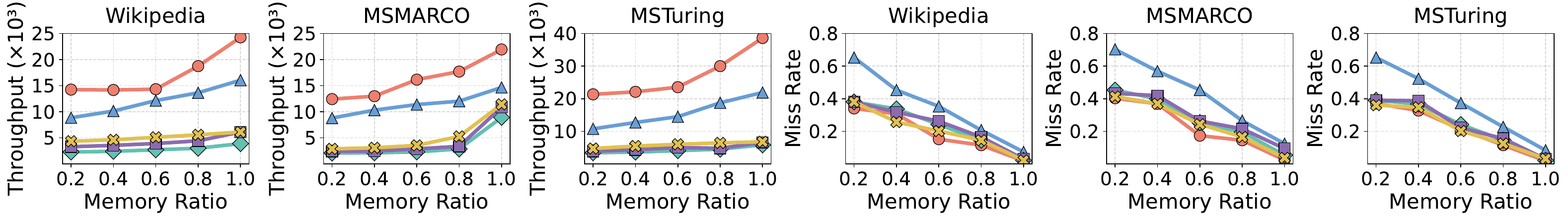}
  \vspace{-2em}
  \caption{Impact of different GPU memory ratios on update throughput and miss rate.}
  \label{fig:gpu_ratio_comparison}
 \vspace{-1.2em}
\end{figure*}

\textbf{Effectiveness of Adaptive Vector Caching.} 
We conduct two complementary experiments to evaluate our caching strategy with four alternatives:
\emph{SVFusion w/o WAVP}, which always computes distances on CPU for cache misses without transferring vectors to GPU; \emph{LRU/LFU/LRFU}, which replace our WAVP mechanism with traditional caching strategies.

\underline{Replacement Strategies.}
We first evaluate the effectiveness of our workload-aware vector placement (WAVP) strategy under the \emph{SlidingWindow} workload.
Figure~\ref{fig:replacement_comparison} demonstrates that SVFusion with WAVP achieves the highest throughput and lowest latency, outperforming SVFusion (w/o WAVP), LRFU, LFU, and LRU by up to 7.2$\times$ in throughput and 5.1$\times$ in latency reduction.
These gains arise because traditional cache policies overlook workload dynamics, causing excessive vector replacements and costly CPU–GPU data transfers.
By adaptively modeling temporal and structural access patterns, WAVP minimizes unnecessary migrations and maximizes GPU utilization, achieving both high efficiency and stable performance across varying workloads.

% Specifically, SVFusion achieves 1.9$\times$, 3.3$\times$, 3.9$\times$, and 7.2$\times$ higher throughput than SVFusion (w/o WAVP), LRFU, LFU, and LRU, respectively.

% This significant performance gap reveals a critical insight: traditional cache strategies incur substantial transfer overhead due to frequent vector replacement, while always handling missing vectors on CPU underutilizes GPU's computational capacity.
% Our adaptive policy strikes a better trade-off between data locality and transfer cost, delivering both high efficiency and performance stability.

\underline{Memory Constraints.} 
% To evaluate the robustness of our caching strategy under varying memory constraints, we fix the dataset size and vary the available GPU memory from 20\% to 100\%.
To evaluate the robustness of our caching strategy under varying GPU memory, we use the same dataset while scaling available GPU memory from 20\% to 100\%.
As shown in Figure~\ref{fig:gpu_ratio_comparison}, our method consistently outperforms all baselines across all settings, achieving 1.7$\times$, 3.3$\times$, 4.1$\times$, and 5.6$\times$ higher average throughput than SVFusion (w/o WAVP), LRFU, LFU, and LRU, respectively. 
Notably, even under reduced GPU capacity, which increases cache miss rates, SVFusion maintains stable throughput. This demonstrates that our design effectively overlaps CPU computation with data transfers, hiding memory latency and sustaining high performance under GPU memory constraints.

\begin{figure}[!t]
    \centering
    \captionsetup[subfigure]{skip=2pt} % 设置子图标题与图片的间距
    \begin{subfigure}[t]{0.33\linewidth}
        \centering
        \includegraphics[width=\linewidth]{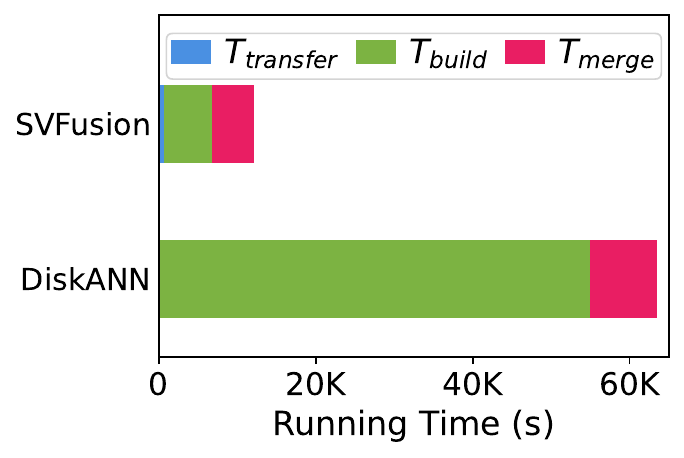}
        \caption{Construction time}
        \label{fig:disk_analysis1}
    \end{subfigure}%
    \hfill%
    \begin{subfigure}[t]{0.33\linewidth}
        \centering
        \includegraphics[width=\linewidth]{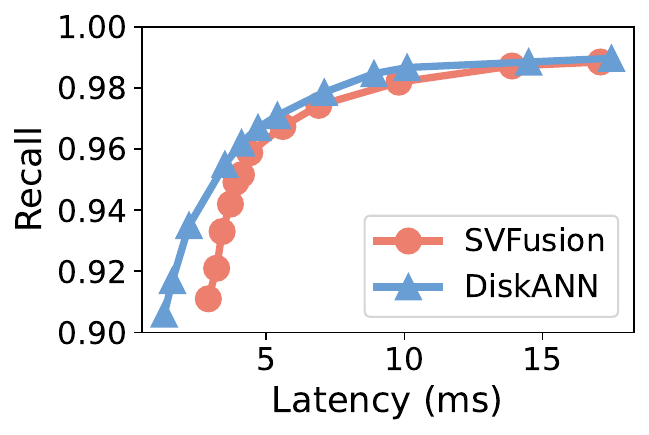}  
        \caption{Recall-Latency} %  
        \label{fig:disk_analysis3}
    \end{subfigure}
    \hfill%
     \begin{subfigure}[t]{0.33\linewidth}
        \centering
        \includegraphics[width=\linewidth]{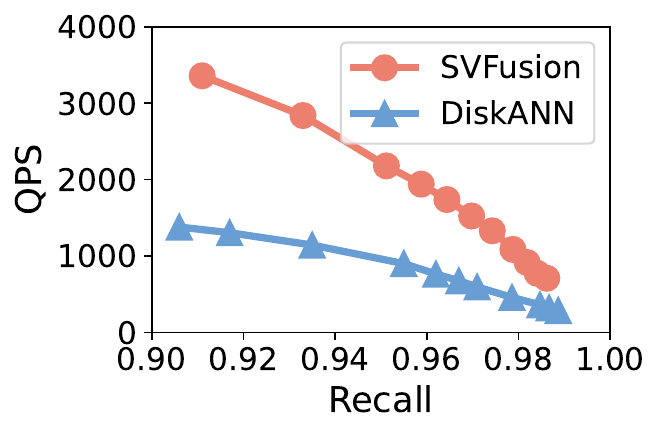}
        \caption{Recall-Throughput}
        \label{fig:disk_analysis2}
    \end{subfigure}%
    \vspace{-3mm}
    \caption{Effect of GPU-CPU-disk framework}
    \label{fig:disk_analysis}
\vspace{-1.5em} 
\end{figure}

\textbf{Effectiveness of Disk tier.}
We evaluate the disk-extended SVFusion framework on the Deep1B dataset~\cite{deep1b}.
As shown in Figure~\ref{fig:disk_analysis1}, SVFusion achieves a 5.26$\times$ faster index construction compared with DiskANN, primarily attributed to a 9.1$\times$ acceleration in GPU-based subgraph building. However, the disk merging stage remains the primary bottleneck in the pipeline.
Figure~\ref{fig:disk_analysis3} and~\ref{fig:disk_analysis2} present the throughput and latency performance across different recall levels.
At high recall targets ($\geq 0.95$), SVFusion achieves latencies that are 0.7–1.6\% higher than those of DiskANN. This minor degradation stems from the increased data partitioning necessitated by limited GPU memory, which slightly weakens cross-partition connectivity.
However, SVFusion outperforms DiskANN in throughput by 2.3$\times$ at comparable recall levels, highlighting the effectiveness of our multi-tier hierarchical index design.

\begin{figure}[t] 
\centering
  \includegraphics[width=0.95\linewidth]{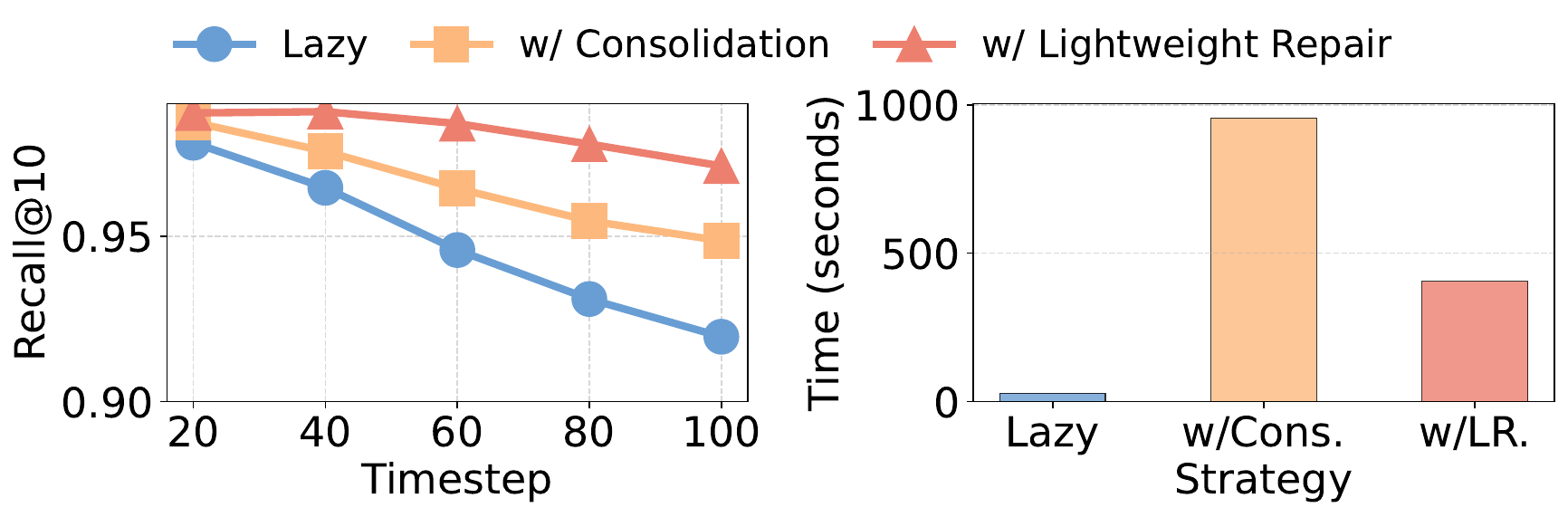}
  \vspace{-1.2em}
  \caption{Effect of deletion strategies.}
  \label{fig:deletion_ablation}
\vspace{-1em}
\end{figure}

\textbf{Effect of Deletion Strategies.}
We compare three approaches: lazy deletion alone, lazy deletion with global consolidation, and our method combining both with localized repair. 
As shown in Figure~\ref{fig:deletion_ablation}, our method achieves 5.2\% and 2.3\% higher recall than the two methods, respectively, while reducing overhead by 57.6\% compared to global consolidation alone.
% This confirms our localized repair strategy effectively preserves index quality with minor performance impacts.

\begin{figure}[!t]
    \centering

    \begin{minipage}[t]{0.49\linewidth}
        \centering
        \includegraphics[width=\linewidth]{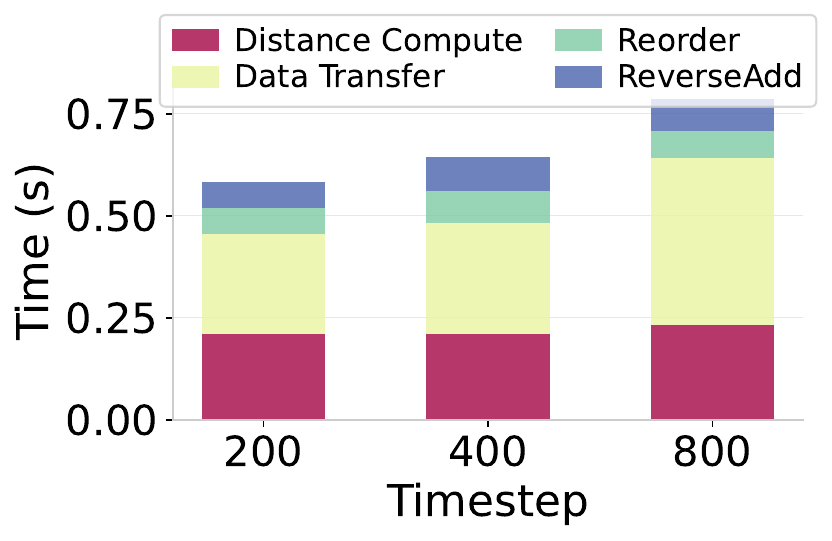}
        \vspace{-2em}
        \caption{Perf. Breakdown.}
        \label{fig:breakdown}
    \end{minipage}
    \hfill
    \begin{minipage}[t]{0.49\linewidth}
        \centering
        %\raisebox{1ex}
        {\includegraphics[width=\linewidth]{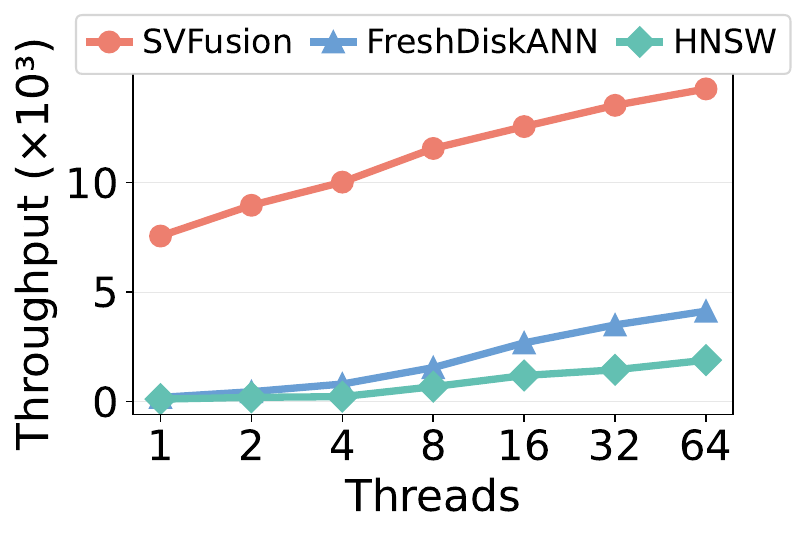}}
        \vspace{-2em}
        \caption{Effect of \#Threads.}
        \label{fig:scalability}
    \end{minipage}

\vspace{-1em}
\end{figure}

\begin{figure}[!t]
    \centering

    \begin{minipage}[t]{0.49\linewidth}
        \centering
        \includegraphics[width=\linewidth]{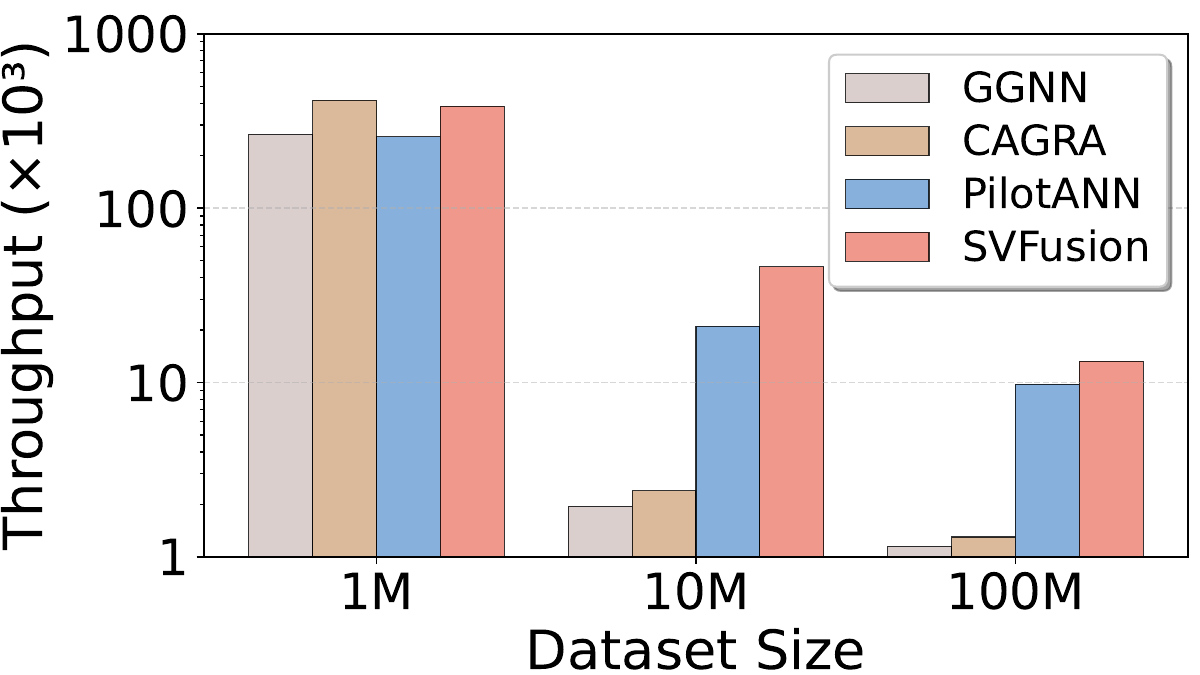}
        \vspace{-2em}
        \caption{GPU methods.}
        \label{fig:gpu_comparison}
    \end{minipage}
    \hfill
    \begin{minipage}[t]{0.49\linewidth}
        \centering
        %\raisebox{1ex}
        {\includegraphics[width=\linewidth]{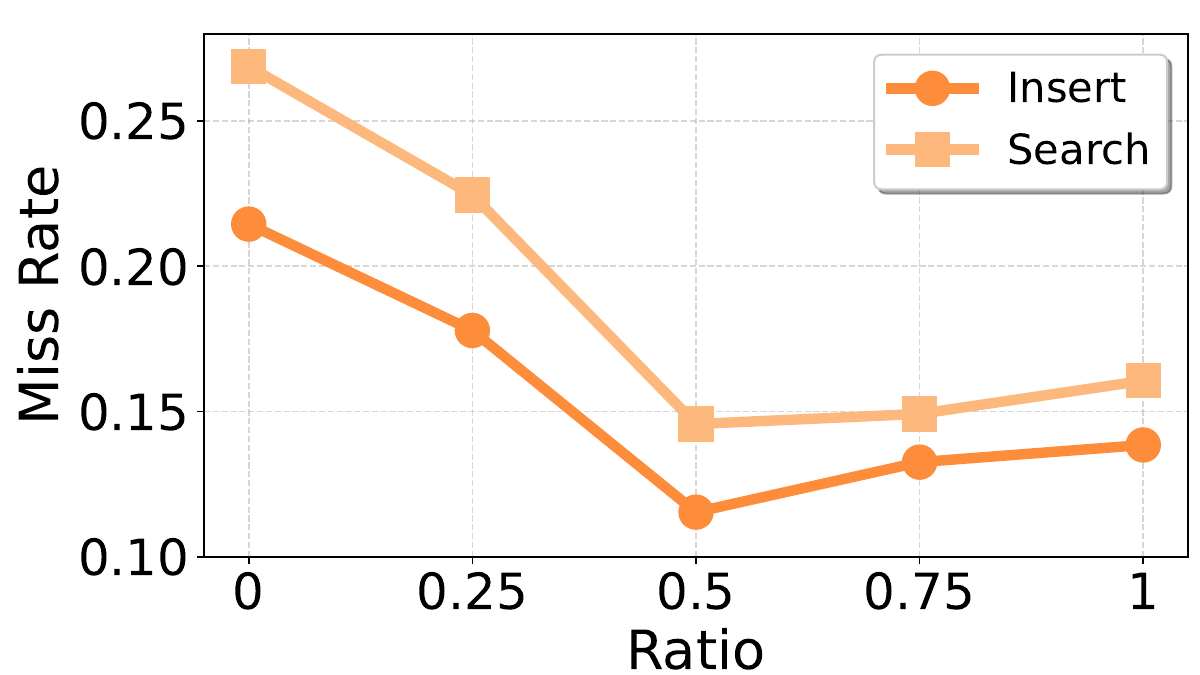}}
        \vspace{-2em}
        \caption{Pred. Param.}
        \label{fig:score_ablation}
    \end{minipage}

\vspace{-1.5em}
\end{figure}

\textbf{Performance Breakdown and Scalability.}
We break down the time cost of insertion of SVFusion into key components (Figure~\ref{fig:breakdown}): \emph{data transfer} (45.4\%), \emph{distance computation} (33.6\%), \emph{heuristic reordering} (10.3\%), and \emph{reverse add} (10.6\%). This highlights two insights: (1) despite GPU acceleration, CPU–GPU data movement remains the dominant bottleneck as index size grows; (2) distance computation remains stable, demonstrating the effectiveness of our pipelined CPU–GPU parallelism. 
We also evaluate scalability by varying thread counts from 1 to 64 (Figure~\ref{fig:scalability}). While all methods show throughput improvements with more threads, the gains diminish beyond 16 threads, with an average of only 1.24$\times$.

\textbf{Effectiveness of CPU-GPU vector search.} Figure~\ref{fig:gpu_comparison} illustrates the CPU-GPU search performance on MSMARCO across different dataset scales under Recall@10=0.9. For datasets that fit within GPU memory, SVFusion achieves comparable performance to CAGRA with slightly lower throughput due to the overhead of maintaining CPU-side data structures. When the dataset exceeds GPU memory, CAGRA and GGNN experience significant performance degradation due to inefficient CPU-GPU data transfers through unified virtual memory (UVM).
Compared to PilotANN, SVFusion achieves 1.5$\times$-2.2$\times$ speedup across all scales due to PilotANN's two-stage design requiring CPU-side refinement.
This demonstrates that our CPU-GPU co-processing approach can efficiently handle large-scale vector search without data compression.

\textbf{Effectiveness of Consistency Guarantee.}
To validate our concurrency control protocol in $\S\ref{sec:cc_update}$, we design a stress test where each batch contains an interleaved mix of 50\% insertions and 50\% search queries, with batch sizes of 10 vectors and request rates ranging from 500 to 10K QPS.
Recall@1 is used as the primary metric, as a correct system should always return the newly inserted vector as the nearest neighbor.
As shown in Table~\ref{tab:consistency_ablation}, SVFusion with synchronization maintains stable Recall@1 at 0.96 across all request rates, confirming strong read-after-write consistency. 
In contrast, the \emph{no-synchronization} variant suffers severe degradation, dropping from 0.96 at 500 QPS to 0.18 at 10K QPS (an 81\% decline), as searches increasingly observe partially updated or inconsistent graph states.
Although enforcing synchronization increases p99 latency from 7.4ms to 33.2ms at the highest load, this overhead is bounded and justified, as it guarantees correctness under extreme streaming workloads. These results demonstrate that our fine-grained, multi-tier coordination protocol effectively balances consistency and performance in high-throughput SANNS scenarios.

\begin{table}[t]
\centering
\caption{Impact of consistency guarantee.}
\label{tab:consistency_ablation}
\vspace{-1em}
\small
\begin{tabular}{l|cccc|cccc}
\toprule
$\backslash$ \textbf{QPS} & \multicolumn{4}{c|}{\textbf{Recall@1}} & \multicolumn{4}{c}{\textbf{p99 Latency (ms)}} \\
& 500 & 2K & 5K & 10K & 500 & 2K & 5K & 10K \\
\midrule
w/ sync  & 0.96 & 0.96 & 0.96 & 0.96 & 4.3  & 6.7 & 10.8 & 33.2\\
w/o sync  & 0.96 & 0.74 & 0.51 & 0.18 & 4.3  & 4.8 & 5.5  & 7.4\\
\bottomrule
\end{tabular}
\vspace{-1.5em}
\end{table}

\subsection{Parameter Sensitivity}
\label{sec:exp_paras}

\textbf{Prediction Parameters.}
We evaluate how the parameters in our prediction function affect miss rates by varying the ratio $\frac{\alpha}{\alpha+\beta}$, where $0$ represents prediction solely on graph topology and $1$ indicates prediction based on recent accesses. Figure~\ref{fig:score_ablation} shows that higher weights for recent accesses generally reduce miss rates, with optimal performance achieved at a ratio around 0.6.

\begin{figure}[t] 
\centering
\includegraphics[width=\linewidth]{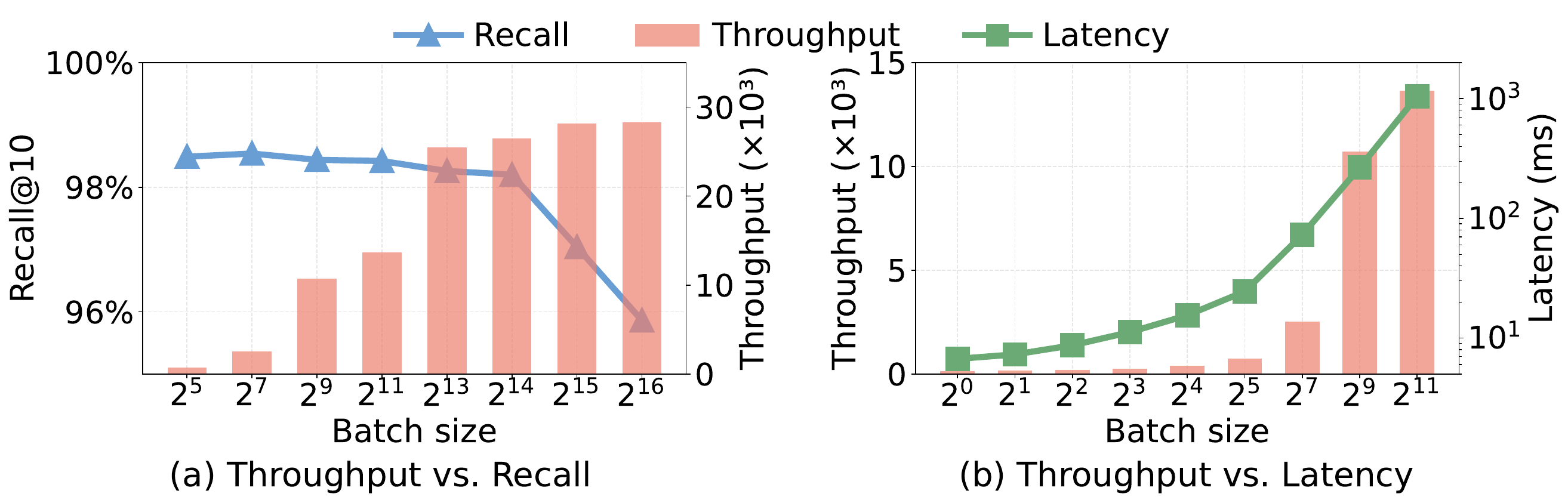}
  \vspace{-2.5em}
  \caption{Impact of varying batch size of updates.}
  \label{fig:batchsize_analysis}
\vspace{-1em}
\end{figure}

\textbf{Batch Size.} 
We analyze the effect of batch size on SVFusion’s performance in Figure~\ref{fig:batchsize_analysis}.
Larger batches substantially increase throughput but degrade recall beyond $2^{13}$ due to delayed graph updates.
Latency also grows exponentially, exceeding one second at $2^{11}$, reflecting the classic latency–throughput trade-off.
SVFusion supports adaptive batch sizing to dynamically balance throughput and query latency for streaming ANNS workloads.

\section{Related Work}
\label{sec:related_work}
\textbf{Vector Indexes.} ANNS is a classical problem with a large body of research work. 
Partition-based methods~\cite{LiteHST,GTI,IVF,PQ,RaBitQ,gao2025practical,mohoney2025quake,hakes} use tree structures and quantization to organize vectors into spatial partitions but suffer from the curse of dimensionality. 
% Partition-based methods leverage tree structures~\cite{LiteHST,GTI}, and quantization techniques~\cite{IVF,PQ,RaBitQ,gao2025practical} to organize vectors into spatial partitions but suffer from the curse of dimensionality. 
Hash-based approaches~\cite{DB-LSH,PM-LSH,Hash-1,gan2012locality} project high-dimensional vectors into lower-dimensional hash buckets to accelerate similarity search. Graph-based methods~\cite{diskann,NSG,elpis,tao-MNG,SeRF,yang2024revisiting} construct proximity graphs to enable efficient vector search. 
% where nodes represent vectors and edges connect similar vectors, enabling efficient navigation during search. 
However, most techniques target static datasets and perform poorly with frequent updates.

\textbf{Streaming Vector Search.}
Recent research has focused on enabling dynamic updates in vector search systems to support streaming scenarios. FreshDiskANN~\cite{freshdiskann} maintains new vectors in memory before periodically merging them into a disk-resident graph. SPFresh~\cite{SPFresh} implements a cluster-based approach with lightweight incremental rebalancing that adjusts partitions as data distribution changes. 
Other approaches~\cite{MN-RU,RTAMS-GANNS,IPDiskANN} focus on maintaining graph connectivity or parallel acceleration.
Despite these advances, existing systems face significant limitations: they either support only limited dataset sizes, handle specific operations (insertions), or exhibit degraded search quality in high-dimensional spaces.
% Other approaches include MN-RU~\cite{MN-RU}, which introduces link adjustment algorithms for maintaining graph connectivity during insertions, and RTAMS-GANNS~\cite{RTAMS-GANNS} supporting real-time vector insertion through multi-stream parallel execution.

\textbf{GPU-accelerated ANNS.}
Leveraging GPUs for vector search has been extensively explored to address computational demands at scale.
Faiss~\cite{faiss} and Raft~\cite{Raft} pioneered CPU-to-GPU migration, while SONG~\cite{SONG}, GANNS~\cite{GANNS}, and CAGRA~\cite{CAGRA} improved GPU parallelism through algorithmic and kernel optimizations.
% Pioneer works like Faiss~\cite{faiss} and Raft~\cite{Raft} first migrated vector retrieval from CPU to GPU.
% SONG~\cite{SONG} reformulated graph-based search to better utilize GPU parallelism, while GANNS~\cite{GANNS} and CAGRA~\cite{CAGRA} have further improved GPU-based algorithms with specialized kernel design.
For system design optimization,~\cite{souza2021online,BANG,RUMMY,PilotANN,GGNN} converged on integrating heterogeneous CPU-GPU scheduling, distributing data between GPU and CPU, and establishing periodic synchronization mechanisms.
However, these methods are designed for static datasets where indexes remain fixed during querying. 
% RTAMS-GANNS~\cite{RTAMS-GANNS} supports real-time insertions but cannot scale to large datasets.
% Thus, efficient dynamic index update on GPUs, particularly for graph-based methods at scale, remains an open challenge.
% The recent RTAMS-GANNS~\cite{RTAMS-GANNS} takes a step toward supporting real-time insertions on GPUs, but it is unable to handle large-scale datasets. 
% Thus, efficient dynamic index updates on GPUs, particularly for graph-based methods at scale, remains an open challenge.
\section{Conclusion}
We presented SVFusion, a GPU-CPU-disk collaborative framework for large-scale and real-time vector search. It introduces a hierarchical vector index spanning memory tiers, enabling efficient data distribution across heterogeneous hardware. We employ an adaptive caching mechanism that optimizes data placement by considering both access patterns and graph structure relationships. SVFusion further incorporates a 
concurrency control protocol that ensures data consistency while maintaining high throughput and low latency.
Experimental results demonstrate that SVFusion achieves up to 71.8$\times$ higher throughput and 1.3$\times$ to 50.7$\times$ lower latency while maintaining superior recall compared to state-of-the-art methods across various streaming workloads.

% \begin{acks}
% This work was supported by the ``Pioneer'' R\&D Program of Zhejiang (2025C01001), CCF-Huawei Populus Grove Fund (Grant No. CCF-HuaweiDB202408), the National Nature Science Foundation of China (62572434), the Fundamental Research Funds for the Central Universities (226-2024-00145).
% % This work was supported by the [...] Research Fund of [...] (Number [...]). Additional funding was provided by [...] and [...]. We also thank [...] for contributing [...].
% \end{acks}

%\clearpage

\bibliographystyle{ACM-Reference-Format}
\bibliography{main}

@misc{Raft,  
  title = {Rapidsai/raft: RAFT contains fundamental widely-used algorithms and primitives for data science, Graph and machine learning.},  
  year = {2022},  
  howpublished = {\url{https://github. com/rapidsai/raft}}, 
}

@misc{Gemini, 
  title = {Personalized help from Gemini.},  
  year = {2025},  
  howpublished = {\url{https://gemini.google/overview/personalization}}, 
}

@misc{letta,  
  title = {The Platform for Building Stateful Agents.},  
  year = {2025},  
  howpublished = {\url{https://www.letta.com/}}, 
}

@misc{CagraIssue,  
  title = {updating an index},  
  year = {2024},  
  howpublished = {\url{https://github.com/rapidsai/cuvs/issues/295}}, 
}

@misc{msturing,  
  title = {Microsoft Turing-ANNS-1B},  
  year = {2024},  
  howpublished = {\url{https://learning2hash.github.io/publications/microsoftturinganns1B/}}, 
}

@misc{wikipedia,  
  title = {wikimedia/wikipedia},  
  year = {2024},  
  howpublished = {\url{https://huggingface.co/datasets/wikimedia/wikipedia}}, 
}

@misc{msmarco,  
  title = {microsoft/ms\_marco},  
  year = {2024},  
  howpublished = {\url{https://huggingface.co/datasets/microsoft/ms_marco}}, 
}

@misc{cuda_guide,  
  title = {CUDA C++ Best Practices Guide},  
  year = {2025},  
  howpublished = {\url{https://docs.nvidia.com/cuda/cuda-c-best-practices-guide/index.html}}, 
}

@misc{OpenaiEmbedding,  
  title = {New embedding models and API updates},  
  year = {2024},  
  howpublished = {\url{https://openai.com/index/new-embedding-models-and-api-updates/}}, 
}

@misc{cuvs,  
  title = {cuVS: Vector Search and Clustering on the GPU},  
  year = {2024},  
  howpublished = {\url{https://github.com/rapidsai/cuvs/tree/v24.12.00}}, 
}

@misc{AlloyDB,  
  title = {ScaNN for AlloyDB: The first PostgreSQL vector search index that works well from millions to billion of vectors},  
  year = {2025},  
  howpublished = {\url{https://cloud.google.com/blog/products/databases/how-scann-for-alloydb-vector-search-compares-to-pgvector-hnsw}}, 
}

@article{MemGPT,
  author       = {Charles Packer and
                  Vivian Fang and
                  Shishir G. Patil and
                  Kevin Lin and
                  Sarah Wooders and
                  Joseph E. Gonzalez},
  title        = {MemGPT: Towards LLMs as Operating Systems},
  journal      = {CoRR},
  volume       = {abs/2310.08560},
  year         = {2023},
}

@article{RAGCache,
  author       = {Chao Jin and
                  Zili Zhang and
                  Xuanlin Jiang and
                  Fangyue Liu and
                  Xin Liu and
                  Xuanzhe Liu and
                  Xin Jin},
  title        = {RAGCache: Efficient Knowledge Caching for Retrieval-Augmented Generation},
  journal      = {CoRR},
  volume       = {abs/2404.12457},
  year         = {2024},
}

@inproceedings{facebook-search,
  author       = {Jui{-}Ting Huang and
                  Ashish Sharma and
                  Shuying Sun and
                  Li Xia and
                  David Zhang and
                  Philip Pronin and
                  Janani Padmanabhan and
                  Giuseppe Ottaviano and
                  Linjun Yang},
  title        = {Embedding-based Retrieval in Facebook Search},
  booktitle    = {Proceedings of the ACM International Conference on Knowledge Discovery and Data Mining (KDD)},
  pages        = {2553--2561},
  year         = {2020},
}

@inproceedings{bing-search,
  author       = {Jianjin Zhang and
                  Zheng Liu and
                  Weihao Han and
                  Shitao Xiao and
                  Ruicheng Zheng and
                  Yingxia Shao and
                  Hao Sun and
                  Hanqing Zhu and
                  Premkumar Srinivasan and
                  Weiwei Deng and
                  Qi Zhang and
                  Xing Xie},
  title        = {Uni-Retriever: Towards Learning the Unified Embedding Based Retriever
                  in Bing Sponsored Search},
  booktitle    = {Proceedings of the ACM International Conference on Knowledge Discovery and Data Mining (KDD)},
  pages        = {4493--4501},
  year         = {2022}
}

@inproceedings{RAG,
  author       = {Patrick Lewis and
                  Ethan Perez and
                  Aleksandra Piktus and
                  Fabio Petroni and
                  Vladimir Karpukhin and
                  Naman Goyal and
                  Heinrich K{\"{u}}ttler and
                  Mike Lewis and
                  Wen{-}tau Yih and
                  Tim Rockt{\"{a}}schel and
                  Sebastian Riedel and
                  Douwe Kiela},
  title        = {Retrieval-Augmented Generation for Knowledge-Intensive {NLP} Tasks},
  booktitle    = {Advances in Neural Information Processing Systems},
  year         = {2020},
}

@article{StonebrakerP24,
  author       = {Michael Stonebraker and
                  Andrew Pavlo},
  title        = {What Goes Around Comes Around... And Around..},
  journal      = {{SIGMOD} Rec.},
  volume       = {53},
  number       = {2},
  pages        = {21--37},
  year         = {2024},
}

@article{PQ,
  author       = {Herv{\'{e}} J{\'{e}}gou and
                  Matthijs Douze and
                  Cordelia Schmid},
  title        = {Product Quantization for Nearest Neighbor Search},
  journal      = {IEEE Transactions on Pattern Analysis and Machine Intelligence (TPAMI).},
  volume       = {33},
  number       = {1},
  pages        = {117--128},
  year         = {2011},
}

@article{RaBitQ,
  author       = {Jianyang Gao and
                  Cheng Long},
  title        = {RaBitQ: Quantizing High-Dimensional Vectors with a Theoretical Error
                  Bound for Approximate Nearest Neighbor Search},
  journal      = {Proc. {ACM} Manag. Data},
  volume       = {2},
  number       = {3},
  pages        = {167},
  year         = {2024},
}

@inproceedings{Vearch,
  author       = {Jie Li and
                  Haifeng Liu and
                  Chuanghua Gui and
                  Jianyu Chen and
                  Zhenyuan Ni and
                  Ning Wang and
                  Yuan Chen},
  title        = {The Design and Implementation of a Real Time Visual Search System
                  on {JD} E-commerce Platform},
  booktitle    = {Proceedings of the International Middleware Conference},
  pages        = {9--16},
  year         = {2018},
}

@inproceedings{SPFresh,
  author       = {Yuming Xu and
                  Hengyu Liang and
                  Jin Li and
                  Shuotao Xu and
                  Qi Chen and
                  Qianxi Zhang and
                  Cheng Li and
                  Ziyue Yang and
                  Fan Yang and
                  Yuqing Yang and
                  Peng Cheng and
                  Mao Yang},
  title        = {SPFresh: Incremental In-Place Update for Billion-Scale Vector Search},
  booktitle    = {Proceedings of  the 29th Symposium on Operating Systems Principles (SOSP).},
  pages        = {545--561},
  year         = {2023}
}

@article{LiteHST,
  author       = {Yuxiang Zeng and
                  Yongxin Tong and
                  Lei Chen},
  title        = {LiteHST: {A} Tree Embedding based Method for Similarity Search},
  journal      = {Proc. {ACM} Manag. Data},
  volume       = {1},
  number       = {1},
  pages        = {35:1--35:26},
  year         = {2023},
}

@article{GTI,
  title={GTI: Graph-Based Tree Index with Logarithm Updates for Nearest Neighbor Search in High-Dimensional Spaces},
  author={Ma, Ruiyao and Zhu, Yifan and Zheng, Baihua and Chen, Lu and Ge, Congcong and Gao, Yunjun},
  journal={Proceedings of the VLDB Endowment},
  volume={18},
  number={4},
  pages={986--999},
  year={2024},
}

@article{IVF,
  title={The inverted multi-index},
  author={Babenko, Artem and Lempitsky, Victor},
  journal={IEEE Transactions on Pattern Analysis and Machine Intelligence (TPAMI).},
  volume={37},
  number={6},
  pages={1247--1260},
  year={2014},
  publisher={IEEE}
}

@article{DB-LSH,
  title={DB-LSH 2.0: Locality-sensitive hashing with query-based dynamic bucketing},
  author={Tian, Yao and Zhao, Xi and Zhou, Xiaofang},
  journal={IEEE Transactions on Knowledge and Data Engineering},
  volume={36},
  number={3},
  pages={1000--1015},
  year={2023}
}

@article{PM-LSH,
  title={PM-LSH: A fast and accurate LSH framework for high-dimensional approximate NN search},
  author={Zheng, Bolong and Xi, Zhao and Weng, Lianggui and Hung, Nguyen Quoc Viet and Liu, Hang and Jensen, Christian S},
  journal={Proceedings of the VLDB Endowment},
  volume={13},
  number={5},
  pages={643--655},
  year={2020}
}

@article{Hash-1,
  author       = {Qiang Huang and
                  Jianlin Feng and
                  Yikai Zhang and
                  Qiong Fang and
                  Wilfred Ng},
  title        = {Query-Aware Locality-Sensitive Hashing for Approximate Nearest Neighbor
                  Search},
  journal      = {Proc. {VLDB} Endow.},
  volume       = {9},
  number       = {1},
  pages        = {1--12},
  year         = {2015}
}

@inproceedings{gan2012locality,
  title={Locality-sensitive hashing scheme based on dynamic collision counting},
  author={Gan, Junhao and Feng, Jianlin and Fang, Qiong and Ng, Wilfred},
  booktitle={Proceedings of the International Conference on Management of Data (SIGMOD).},
  pages={541--552},
  year={2012}
}

@article{hnsw,
  author       = {Yury A. Malkov and
                  Dmitry A. Yashunin},
  title        = {Efficient and Robust Approximate Nearest Neighbor Search Using Hierarchical
                  Navigable Small World Graphs},
  journal      = {IEEE Transactions on Pattern Analysis and Machine Intelligence (TPAMI).},
  volume       = {42},
  number       = {4},
  pages        = {824--836},
  year         = {2020},
}

@article{freshdiskann,
  title={Freshdiskann: A fast and accurate graph-based ann index for streaming similarity search},
  author={Singh, Aditi and Subramanya, Suhas Jayaram and Krishnaswamy, Ravishankar and Simhadri, Harsha Vardhan},
  journal={arXiv preprint arXiv:2105.09613},
  year={2021}
}

@article{diskann,
  title={Diskann: Fast accurate billion-point nearest neighbor search on a single node},
  author={Jayaram Subramanya, Suhas and Devvrit, Fnu and Simhadri, Harsha Vardhan and Krishnawamy, Ravishankar and Kadekodi, Rohan},
  journal={Advances in neural information processing Systems},
  volume={32},
  year={2019}
}

@article{NSG,
  title={Fast Approximate Nearest Neighbor Search With The Navigating Spreading-out Graph},
  author={Fu, Cong and Xiang, Chao and Wang, Changxu and Cai, Deng},
  journal={Proceedings of the VLDB Endowment},
  volume={12},
  number={5},
  pages={461--474},
  year={2019}
}

@article{elpis,
  title={Elpis: Graph-based similarity search for scalable data science},
  author={Azizi, Ilias and Echihabi, Karima and Palpanas, Themis},
  journal={Proceedings of the VLDB Endowment},
  volume={16},
  number={6},
  pages={1548--1559},
  year={2023},
  publisher={VLDB Endowment}
}

@article{tao-MNG,
  title={Efficient approximate nearest neighbor search in multi-dimensional databases},
  author={Peng, Y and Choi, B and Chan, TN and Yang, J and Xu, J},
  journal={Proceedings of the International Conference on Management of Data (SIGMOD).},
  year={2023},
  publisher={ACM}
}

@article{serf,
  title={SeRF: segment graph for range-filtering approximate nearest neighbor search},
  author={Zuo, Chaoji and Qiao, Miao and Zhou, Wenchao and Li, Feifei and Deng, Dong},
  journal={Proceedings of the ACM on Management of Data (SIGMOD)},
  volume={2},
  number={1},
  pages={1--26},
  year={2024}
}

@article{MN-RU,
  title={Enhancing HNSW Index for Real-Time Updates: Addressing Unreachable Points and Performance Degradation},
  author={Xiao, Wentao and Zhan, Yueyang and Xi, Rui and Hou, Mengshu and Liao, Jianming},
  journal={arXiv preprint arXiv:2407.07871},
  year={2024}
}

@inproceedings{RTAMS-GANNS,
  title={A Real-Time Adaptive Multi-Stream GPU System for Online Approximate Nearest Neighborhood Search},
  author={Sun, Yiping and Shi, Yang and Du, Jiaolong},
  booktitle={Proceedings of the 33rd ACM International Conference on Information and Knowledge Management},
  pages={4906--4913},
  year={2024}
}

@article{faiss,
  title={Billion-scale similarity search with GPUs},
  author={Johnson, Jeff and Douze, Matthijs and J{\'e}gou, Herv{\'e}},
  journal={IEEE Transactions on Big Data},
  volume={7},
  number={3},
  pages={535--547},
  year={2019},
  publisher={IEEE}
}

@inproceedings{CAGRA,
  author       = {Hiroyuki Ootomo and
                  Akira Naruse and
                  Corey Nolet and
                  Ray Wang and
                  Tamas Feher and
                  Yong Wang},
  title        = {{CAGRA:} Highly Parallel Graph Construction and Approximate Nearest Neighbor Search for GPUs},
  booktitle    = {{IEEE} International Conference on Data Engineering (ICDE).},
  pages        = {4236--4247},
  year         = {2024},
}

@inproceedings{FusionANNS,
  author       = {Bing Tian and
                  Haikun Liu and
                  Yuhang Tang and
                  Shihai Xiao and
                  Zhuohui Duan and
                  Xiaofei Liao and
                  Hai Jin and
                  Xuecang Zhang and
                  Junhua Zhu and
                  Yu Zhang},
  title        = {Towards High-throughput and Low-latency Billion-scale Vector Search via {CPU/GPU} Collaborative Filtering and Re-ranking},
  booktitle    = {{USENIX} Conference on File and Storage Technologies (FAST).},
  pages        = {171--185},
  year         = {2025},
}

@inproceedings{RUMMY,
  author       = {Zili Zhang and
                  Fangyue Liu and
                  Gang Huang and
                  Xuanzhe Liu and
                  Xin Jin},
  title        = {Fast Vector Query Processing for Large Datasets Beyond {GPU} Memory
                  with Reordered Pipelining},
  booktitle    = {{USENIX} Symposium on Networked Systems Design and Implementation (NSDI)},
  pages        = {23--40},
  year         = {2024},
}

@inproceedings{SONG,
  author       = {Weijie Zhao and
                  Shulong Tan and
                  Ping Li},
  title        = {{SONG:} Approximate Nearest Neighbor Search on {GPU}},
  booktitle    = {{IEEE} International Conference on Data Engineering (ICDE)},
  pages        = {1033--1044},
  year         = {2020}
}

@article{BANG,
  title={BANG: Billion-Scale Approximate Nearest Neighbor Search using a Single GPU},
  author={Khan, Saim and Singh, Somesh and Simhadri, Harsha Vardhan and Vedurada, Jyothi and others},
  journal={arXiv preprint arXiv:2401.11324},
  year={2024}
}

@article{souza2021online,
  title={Online multimedia retrieval on CPU--GPU platforms with adaptive work partition},
  author={Souza, Rafael and Fernandes, Andr{\'e} and Teixeira, Thiago SFX and Teodoro, George and Ferreira, Renato},
  journal={Journal of Parallel and Distributed Computing},
  volume={148},
  pages={31--45},
  year={2021},
  publisher={Elsevier}
}

@inproceedings{GANNS,
  title={GPU-accelerated proximity graph approximate nearest Neighbor search and construction},
  author={Yu, Yuanhang and Wen, Dong and Zhang, Ying and Qin, Lu and Zhang, Wenjie and Lin, Xuemin},
  booktitle={2022 IEEE 38th International Conference on Data Engineering (ICDE)},
  pages={552--564},
  year={2022},
  organization={IEEE}
}

@article{GGNN,
  author       = {Fabian Groh and
                  Lukas Ruppert and
                  Patrick Wieschollek and
                  Hendrik P. A. Lensch},
  title        = {{GGNN:} Graph-Based {GPU} Nearest Neighbor Search},
  journal      = {{IEEE} Trans. Big Data},
  volume       = {9},
  number       = {1},
  pages        = {267--279},
  year         = {2023},
}

@article{manu,
author = {Guo, Rentong and Luan, Xiaofan and Xiang, Long and Yan, Xiao and Yi, Xiaomeng and Luo, Jigao and Cheng, Qianya and Xu, Weizhi and Luo, Jiarui and Liu, Frank and Cao, Zhenshan and Qiao, Yanliang and Wang, Ting and Tang, Bo and Xie, Charles},
title = {Manu: a cloud native vector database management system},
year = {2022},
volume = {15},
number = {12},
journal = {Proc. VLDB Endow.},
month = aug,
pages = {3548–3561},
numpages = {14}
}

@article{wang21graphanns,
author = {Wang, Mengzhao and Xu, Xiaoliang and Yue, Qiang and Wang, Yuxiang},
title = {A comprehensive survey and experimental comparison of graph-based approximate nearest neighbor search},
year = {2021},
issue_date = {July 2021},
publisher = {VLDB Endowment},
volume = {14},
number = {11},
journal = {Proc. VLDB Endow.},
month = jul,
pages = {1964–1978},
numpages = {15}
}

@article{annbenchmarks,
  title={ANN-Benchmarks: A benchmarking tool for approximate nearest neighbor algorithms},
  author={Aum{\"u}ller, Martin and Bernhardsson, Erik and Faithfull, Alexander},
  journal={Information Systems},
  volume={87},
  pages={101374},
  year={2020},
}

@article{SSG,
author = {Fu, Cong and Wang, Changxu and Cai, Deng},
title = {High Dimensional Similarity Search With Satellite System Graph: Efficiency, Scalability, and Unindexed Query Compatibility},
year = {2022},
volume = {44},
number = {8},
journal = {IEEE Trans. Pattern Anal. Mach. Intell. (TPAMI)},
pages = {4139–4150},
numpages = {12}
}

@inproceedings{nns,
  author       = {Kenneth L. Clarkson},
  title        = {An Algorithm for Approximate Closest-Point Queries},
  booktitle    = {Proceedings of the Symposium on Computational Geometry},
  pages        = {160--164},
  year         = {1994},
}

@article{simhadri2024results,
  title={Results of the Big ANN: NeurIPS'23 competition},
  author={Simhadri, Harsha Vardhan and Aum{\"u}ller, Martin and Ingber, Amir and Douze, Matthijs and Williams, George and Manohar, Magdalen Dobson and Baranchuk, Dmitry and Liberty, Edo and Liu, Frank and Landrum, Ben and others},
  journal={arXiv preprint arXiv:2409.17424},
  year={2024}
}

@article{liu2004investigation,
  title={An investigation of practical approximate nearest neighbor algorithms},
  author={Liu, Ting and Moore, Andrew and Yang, Ke and Gray, Alexander},
  journal={NIPS},
  volume={17},
  year={2004}
}

@article{upreti2025cost,
  title={Cost-Effective, Low Latency Vector Search with Azure Cosmos DB},
  author={Upreti, Nitish and Sundaram, Krishnan and Sundar, Hari Sudan and Boshra, Samer and Perumalswamy, Balachandar and Atri, Shivam and Chisholm, Martin and Singh, Revti Raman and Yang, Greg and Pattipaka, Subramanyam and others},
  journal={arXiv preprint arXiv:2505.05885},
  year={2025}
}

@inproceedings{LRU,
author = {O'Neil, Elizabeth J. and O'Neil, Patrick E. and Weikum, Gerhard},
title = {The LRU-K page replacement algorithm for database disk buffering},
year = {1993},
booktitle = {Proceedings of the International Conference on Management of Data (SIGMOD).},
pages = {297–306},
numpages = {10},
}

@article{PilotANN,
  author       = {Yuntao Gui and
                  Peiqi Yin and
                  Xiao Yan and
                  Chaorui Zhang and
                  Weixi Zhang and
                  James Cheng},
  title        = {PilotANN: Memory-Bounded {GPU} Acceleration for Vector Search},
  journal      = {CoRR},
  volume       = {abs/2503.21206},
  year         = {2025}
}

@article{gao2025practical,
  title={Practical and asymptotically optimal quantization of high-dimensional vectors in euclidean space for approximate nearest neighbor search},
  author={Gao, Jianyang and Gou, Yutong and Xu, Yuexuan and Yang, Yongyi and Long, Cheng and Wong, Raymond Chi-Wing},
  journal={Proceedings of the ACM on Management of Data},
  volume={3},
  number={3},
  pages={1--26},
  year={2025}
}

@article{yang2024revisiting,
  author       = {Yang, Shuo and Xie, Jiadong and Liu, Yingfan and Yu, Jeffrey Xu and Gao, Xiyue and Wang, Qianru and Peng, Yanguo and Cui, Jiangtao},
  title        = {Revisiting the index construction of proximity graph-based approximate nearest neighbor search},
  journal      = {Proc. {VLDB} Endow.},
  volume       = {18},
  number       = {6},
  pages        = {1825-1838},
  year         = {2025}
}

@article{IPDiskANN,
  author       = {Haike Xu and
                  Magdalen Dobson Manohar and
                  Philip A. Bernstein and
                  Badrish Chandramouli and
                  Richard Wen and
                  Harsha Vardhan Simhadri},
  title        = {In-Place Updates of a Graph Index for Streaming Approximate Nearest
                  Neighbor Search},
  journal      = {CoRR},
  volume       = {abs/2502.13826},
  year         = {2025}
}

@inproceedings{mohoney2025quake,
author = {Mohoney, Jason and Sarda, Devesh and Tang, Mengze and Chowdhury, Shihabur Rahman and Pacaci, Anil and Ilyas, Ihab F. and Rekatsinas, Theodoros and Venkataraman, Shivaram},
title = {Quake: adaptive indexing for vector search},
year = {2025},
booktitle = {Proceedings of the USENIX Conference on Operating Systems Design and Implementation},
articleno = {9},
numpages = {17},
}

@inproceedings{ParlayANN,
author = {Manohar, Magdalen Dobson and Shen, Zheqi and Blelloch, Guy and Dhulipala, Laxman and Gu, Yan and Simhadri, Harsha Vardhan and Sun, Yihan},
title = {ParlayANN: Scalable and Deterministic Parallel Graph-Based Approximate Nearest Neighbor Search Algorithms},
year = {2024},
booktitle = {Proceedings of the SIGPLAN Annual Symposium on Principles and Practice of Parallel Programming},
pages = {270–285},
numpages = {16},
}

@inproceedings{Iwasaki16,
  author       = {Masajiro Iwasaki},
  title        = {Pruned Bi-directed K-nearest Neighbor Graph for Proximity Search},
  booktitle    = {International Conference on Similarity Search and Applications},
  volume       = {9939},
  pages        = {20--33},
  year         = {2016},
}

@article{ADSampling,
author = {Gao, Jianyang and Long, Cheng},
title = {High-Dimensional Approximate Nearest Neighbor Search: with Reliable and Efficient Distance Comparison Operations},
year = {2023},
volume = {1},
number = {2},
journal = {Proc. ACM Manag. Data},
}

@article{lrfu,
  title={LRFU: A spectrum of policies that subsumes the least recently used and least frequently used policies},
  author={Lee, Donghee and Choi, Jongmoo and Kim, Jong-Hun and Noh, Sam H and Min, Sang Lyul and Cho, Yookun and Kim, Chong Sang},
  journal={IEEE transactions on Computers},
  volume={50},
  number={12},
  pages={1352--1361},
  year={2001}
}

@inproceedings{deep1b,
  title={Efficient indexing of billion-scale datasets of deep descriptors},
  author={Babenko, Artem and Lempitsky, Victor},
  booktitle={Proceedings of the IEEE Conference on Computer Vision and Pattern Recognition},
  pages={2055--2063},
  year={2016}
}

@misc{text2image,  
  title = {Benchmarks for Billion-Scale Similarity Search},  
  author = {Yandex Research},
  year = {2021},  
  howpublished = {\url{https://research.yandex.com/blog/benchmarks-for-billion-scale-similarity-search}}, 
}

@inproceedings{dong2011,
  title={Efficient k-nearest neighbor graph construction for generic similarity measures},
  author={Dong, Wei and Moses, Charikar and Li, Kai},
  booktitle={WWW},
  pages={577--586},
  year={2011}
}

@misc{streaming_rag,  
  title = {Streaming Data to RAG},  
  author = {NVIDIA},
  year = {2025},  
  howpublished = {\url{https://build.nvidia.com/nvidia/streaming-data-to-rag/blueprintcard}}, 
}

@article{rong2021scheduling,
  title={Scheduling massive camera streams to optimize large-scale live video analytics},
  author={Rong, Chenghao and Wang, Jessie Hui and Liu, Juncai and Wang, Jilong and Li, Fenghua and Huang, Xiaolei},
  journal={IEEE/ACM Transactions on Networking},
  volume={30},
  number={2},
  pages={867--880},
  year={2021},
  publisher={IEEE}
}

@misc{nips_2023_streaming,  
  title = {NeurIPS 2023 Streaming Challenge and Beyond},  
  year = {2023},  
  howpublished = {\url{https://github.com/harsha-simhadri/big-ann-benchmarks/tree/main/neurips23/streaming}}, 
}

@article{hakes,
author = {Hu, Guoyu and Cai, Shaofeng and Dinh, Tien Tuan Anh and Xie, Zhongle and Yue, Cong and Chen, Gang and Ooi, Beng Chin},
title = {HAKES: Scalable Vector Database for Embedding Search Service},
year = {2025},
volume = {18},
number = {9},
journal = {Proc. VLDB Endow.},
pages = {3049–3062},
numpages = {14}
}

\end{document}